\documentclass[
  prb,
  twocolumn,
  preprintnumbers,
  floatfix,
  showpacs,
  amsmath,
  amssymb,
]{revtex4-1}

\usepackage{amsmath}
\usepackage{amssymb} 
\usepackage{color}
\usepackage{graphicx}
\usepackage{tabularx}
\usepackage{subfigure}
\usepackage{dcolumn}
\usepackage{times}
  
\newcommand{\Al}[0]{\text{Al}}
\newcommand{\Sb}[0]{\text{Sb}}
\newcommand{\AlSb}[0]{\text{AlSb}}

\newcommand{\bulk}[0]{\text{bulk}}
\newcommand{\VB}[0]{\text{VB}}
\newcommand{\CB}[0]{\text{CB}}
\newcommand{\VBM}[0]{\text{VBM}}

\renewcommand{\deg}{\ensuremath{^\circ}}

\newcommand{\eV}[0]{\text{eV}}
\newcommand{\meV}[0]{\text{meV}}

\newcommand{\cm}[0]{\text{cm}}

\newcommand{\atom}[0]{\text{atom}}

\newcommand{\AAA}[0]{\text{\AA}}

\newcommand{\ntype}[0]{\textit{n}-type }
\newcommand{\ptype}[0]{\textit{p}-type }

\newcommand{\sect}[1]{Sec.~\ref{#1}}
\newcommand{\fig}[1]{Fig.~\ref{#1}}
\newcommand{\eq}[1]{Eq.~\eqref{#1}}
\newcommand{\tab}[1]{Table~\ref{#1}}

\renewcommand{\epsilon}[0]{\varepsilon}

\newlength{\myhgt}

\begin{document}

\preprint{published in Phys. Rev. B {\bf 77}, 165206 (2008) Copyright (2008) by the American Physical Society}


\title{
  Intrinsic point defects in aluminum antimonide
}

\author{Daniel {\AA}berg}
\email{aberg2@llnl.gov}
\author{Paul Erhart}
\author{Andrew J. Williamson}
\author{Vincenzo Lordi}
\email{lordi2@llnl.gov}
\affiliation{
  Lawrence Livermore National Laboratory,
  Livermore, California
}

\begin{abstract}
Calculations within density functional theory on the basis of the local density approximation are carried out to study the properties of intrinsic point defects in aluminum antimonide. Special care is taken to address finite-size effects, band gap error, and symmetry reduction in the defect structures. The correction of the band gap is based on a set of {\it GW} calculations. The most important defects are identified to be the aluminum interstitial $\Al_{i,\Al}^{1+}$, the antimony antisites $\Sb_{\Al}^0$ and $\Sb_{\Al}^{1+}$, and the aluminum vacancy $V_{\Al}^{3-}$. The intrinsic defect and charge carrier concentrations in the impurity-free material are calculated by self-consistently solving the charge neutrality equation. The impurity-free material is found to be \ntype conducting at finite temperatures.
\end{abstract}

\maketitle

\section{Introduction}
\label{sect:introduction}

Aluminum antimonide is receiving much renewed interest for applications ranging from ionizing radiation detection to microelectronics to optoelectronics. For gamma radiation detection, AlSb is particularly promising as a novel material enabling high energy-resolution detection at room temperature due to its indirect band gap of 1.6\,eV, the high atomic number of Sb, and the potentially high electron and hole mobilities of up to several hundred cm$^2$\,V$^{-1}$\,s$^{-1}$ at room temperature.\cite{Armantrout, OwensReview} Improved materials for high resolution, room temperature gamma radiation detectors are  critically needed for applications in nuclear nonproliferation and monitoring, homeland security, and also medical and space imaging applications. Such detectors operate by counting the number of electron-hole pairs created in the semiconductor upon interaction with a gamma ray. Thus, a small band gap is generally desired to maximize the number of generated carriers, increasing the signal and reducing the shot noise. However, if the band gap is too small, excess thermal noise is generated from carriers thermally excited across the gap. At room temperature, a gap of $\sim$1.6\,eV is nearly optimal. In a similar vein, low intrinsic carrier concentrations in the material may be desired to reduce the background signal, another noise source. Furthermore, high energy resolution (counting statistics) is achieved by maximizing the efficiency of charge collection, which requires high carrier mobilities and long carrier lifetimes. An indirect band gap can be advantageous for maximizing carrier lifetimes by quenching radiative recombination. Finally, high atomic number in the material is desired to increase the stopping power for high energy radiation, reducing the required size of the device. Presently, the purity of large single-crystal growths of AlSb limits its performance for radiation detection application.

In microelectronics, AlSb and related alloys containing In, are finding use in advanced field-effect transistor designs that promise higher switching speeds and lower power consumption compared to silicon devices.\cite{TuttleKroemer, Ashley} The material has also found use in high current density, high speed resonant tunnel diodes.\cite{RTD} In optoelectronics, thin films of AlSb are being used in active regions and superlattice structures for claddings in novel type-II infrared cascade lasers.\cite{Meyer, Meyer2, Montpelier} Such lasers, emitting wavelengths around 3--4.3\,$\mu$m, find application, for example, in remote atmospheric chemical sensing. Aluminum antimonide is particularly interesting in these applications because it has a large 2.1\,eV conduction band offset with InAs, which often comprises the other key component in the active layers of these devices.

In the present work, we present a careful analysis of the thermodynamic and electronic properties of intrinsic point defects in AlSb, since the electronic and transport properties of the material can be degraded by detrimental defects. The main goals of this study are to identify the most important intrinsic point defects and to establish the concentrations of defects and charge carriers in thermal equilibrium. This work is part of a larger effort to understand the fundamental microscopic limits of performance of this and other semiconductor materials. Future work will focus on extrinsic impurities in the material, as well as the implications on carrier transport properties.

Our major findings in this work are that the dominant native defects in AlSb are aluminum interstitials, antimony antisites, and aluminum vacancies, dependent on chemical environment and doping. The equilibrium concentration of total native defects near the melting temperature is found to be in the $10^{16}$ to $10^{17}$\,cm$^{-3}$ range, while at lower temperatures, concentrations down to $10^{10}$\,cm$^{-3}$ or lower are expected. The electron chemical potential in pure material is near the middle of the band gap, however the material tends naturally to be slightly \ntype doped by charged aluminum interstitials.

The paper is organized as follows. In \sect{sect:method} we review the thermodynamic formalism we use to derive defect formation energies and concentrations from first principles calculations, and describe the computational details which underlie the present work. In \sect{sect:results} we report the relaxed defect geometries, defect formation energies, and the charge carrier and defect concentrations, which are obtained by self-consistently solving the charge neutrality condition. Finally, in \sect{sect:discussion} we discuss the intrinsic limitations of AlSb on the basis of our results and the relation to experimental data. The Appendix contains a brief derivation of the band gap correction scheme utilized in the present work.

\section{Methodology}
\label{sect:method}

\subsection{Point defect thermodynamics}
\label{sect:method_eform}

A material in thermodynamic equilibrium must contain a certain number of point defects at finite temperature, due to entropy. The temperature dependence of the equilibrium defect concentration can be shown to obey the following relation \cite{AllLid03}
\begin{align}
  c &= c_0 \exp \left( -\frac{\Delta G_D}{k_B T} \right),
  \label{eq:concentration}
\end{align}
where $c_0$ denotes the concentration of possible defect sites, $k_B$ is Boltzmann's constant, and $T$ is absolute temperature. The Gibbs energy of defect formation, $\Delta G_D$, can be split into three distinct terms
\begin{align}
  \Delta G_D &= \Delta E_D + T \Delta S_D + p \Delta V_D.
\end{align}
The formation entropy, $\Delta S_D$, is typically on the order of $1\,k_B$, so the entropy term hardly exceeds 0.1\,eV even at elevated temperatures. In some cases, entropic effects could play a role in stabilizing a defect at high temperatures when the enthalpy differences are small. We do not explicitly treat this effect here, but point out cases in our results where it may be important. The formation volume, $\Delta V_D$, describes the pressure dependence of the Gibbs energy of formation and is typically some small fraction of the atomic volume,  so it can be neglected at ordinary pressures. The most important contribution is the defect formation energy, $\Delta E_D$. In the following, we therefore focus on the formation energy and for simplicity ignore both the entropy and volume terms.

For a binary compound the formation energy for a defect in charge state $q$ is given by\cite{QiaMarCha88, ZhaNor91}
\begin{align}
  \Delta E_D
  &= E_D
  - \frac{1}{2} \left( n_{\Al} + n_{\Sb} \right) \mu_{\AlSb}^{\bulk}
  \nonumber
  \\
  & \quad
  - \frac{1}{2} \left( n_{\Al} - n_{\Sb} \right)
  \left( \mu^{\bulk}_{\Al} - \mu^{\bulk}_{\Sb} \right)
  \nonumber
  \\
  & \quad
  - \frac{1}{2} \left( n_{\Al} - n_{\Sb} \right) \Delta\mu
  + q \left( E_{\VBM} + \mu_e \right),
  \label{eq:eform}
\end{align}
where $E_D$ is the total energy of the system containing the defect, $n_i$ denotes the number of atoms of type $i$, $\mu_i^{\bulk}$ is the chemical potential of component $i$ in its reference state, and we have written the terms using explicit labels for AlSb. Neglecting entropic contributions, the chemical potentials of the reference phases can be replaced by their cohesive energies at 0\,K. The formation energy depends on the chemical environment via the parameter $\Delta\mu$, which describes the variation of the chemical potentials under different conditions. The range of $\Delta\mu$ is constrained by the formation energy of AlSb by $|\Delta\mu|\leq \Delta H_f^\AlSb$, where for the present convention $\Delta\mu=-\Delta H_f^\AlSb$ and $\Delta\mu=+\Delta H_f^\AlSb$ correspond to Al and Sb-rich conditions, respectively. Finally, the formation energy also depends on the electron chemical potential, $\mu_e$, which is measured with respect to the valence band maximum, $E_{\VBM}$.

\subsection{Computational details}

The energy terms in \eq{eq:eform} were calculated using density functional theory (DFT) carried out in the local density approximation (LDA) using the Vienna {\it Ab-initio} Simulation Package (\textsc{vasp}) \cite{KreHaf93, KreHaf94, KreFur96a, KreFur96b} and the projector augmented-wave (PAW) method.\cite{Blo94, KreJou99} Defect formation energies were obtained using supercells of various size containing 32, 64, 128, and 216 atoms. Extrapolation was used to account for finite-size effects as described in detail in \sect{sect:finitesize}. Brillouin zone integrations were performed with $k$-point grids generated using the Monkhorst-Pack scheme.\cite{MonPac76} For the 32 and 64-atom cells, a non-shifted $6\times 6\times 6$ mesh was used, while for the 128-atom cell, a shifted $3\times 3\times 3$ grid was used. For the 216-atom cell, a non-shifted $4\times 4\times 4$ mesh was constructed. The plane wave cutoff energy was set to 300\,eV and Gaussian smearing with a width of 0.1\,eV was used to determine the occupation numbers.  For charged defect calculations, a homogeneous background charge was employed (by omitting the $G=0$ term in the potential) to ensure charge neutrality of the entire cell.

Atomic relaxations were performed to determine the equilibrium structures of the defects, with ionic forces converged to 20\,\meV/\AA\ and all calculations performed at the theoretical equilibrium volume. Relaxations from various randomized initial configurations were performed to avoid high-symmetry local energy minima in the structures. 

\subsection{Finite-size corrections}
\label{sect:finitesize}

\begin{figure}
  \includegraphics[width=0.9\columnwidth]{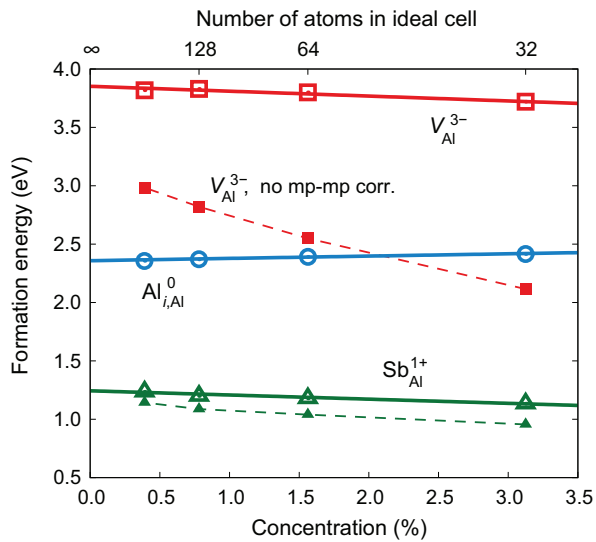}
  \caption{
    (Color online) Illustration of finite-size scaling of formation energies to infinite dilution (concentration $\rightarrow$ 0\%) for three of the most important defects, with no band gap corrections and $\Delta\mu_e=0\,\eV$. The small filled and large open symbols respectively show the data without and with the monopole-monopole correction term. Note that, particularly for the charged aluminum vacancy, the monopole-monopole correction is the dominant size-dependent term. The monopole-monopole correction does not apply for neutral defects.
  }
  \label{fig:finitesize}
\end{figure}

In the supercell approximation there are spurious interactions between defects and their periodic images which lead to systematic errors.\cite{ZhaPerLan04, ErhAlbKle06} For neutral defects the leading error is due to elastic interactions, which cause an overestimation of the formation energy. The strain energy of a point-like inclusion can be derived from linear elasticity theory and can be shown to fall off roughly with $L^{-3}$, where $L$ is the distance between periodic images. \cite{GreDed71, DedPol72} Therefore, the formation energy in the dilute limit ($L \rightarrow \infty$) can be obtained by finite-size scaling with $L^{-3}$, which removes the elastic strain component.

Makov and Payne considered the convergence of the energy of charged systems in periodic systems and proposed a correction on the basis of a multi\-pole ex\-pan\-sion.\cite{MakPay95} The leading term corresponds to the monopole-monopole interaction and scales with $L^{-1}$. This term can be ana\-lytically determined if the static dielectric constant of the medium, $\epsilon$, and the Madelung constant of the Bravais lattice of the supercell, $\alpha$, are known:\cite{MakPay95, LenMozNie02}
\begin{align}
  \Delta E_\text{mp} = -\frac{q^2 \alpha}{2 L \epsilon}.
  \label{eq:MakovPayne}
\end{align}
The next higher order term in the expansion is the monopole-quadrupole interaction which scales as $L^{-3}$. Even higher order terms ($\mathcal{O}(L^{-n})$, $n\geq 5$) are usually small and therefore neglected. In the present work, we have applied the monopole-monopole correction term using the experimental value for the static dielectric constant ($\epsilon = 12$). Then, since both the elastic and the monopole-quadrupole interactions scale with $L^{-3}$, we employed finite-size scaling with $L^{-3}$ to correct for these terms. This extrapolation scheme gave very small extrapolation errors as shown in Tables~\ref{tab:eform1} and \ref{tab:eform2}. The results of the finite-size scaling procedure are illustrated in \fig{fig:finitesize} for the most important defects. Figure~\ref{fig:finitesize} clearly illustrates the effect of the monopole-monopole correction term which tremendously reduces the variation between the supercells. The remaining higher-order variations are very well captured by the $L^{-3}$ finite-size scaling. In addition, the potential alignment correction described in Ref.~\onlinecite{ZhaPerLan04} is implicitly taken into account by our extrapolation scheme.\cite{ErhAlb07a}

\subsection{Band gap corrections}
\label{sect:EG_corr}

The underestimation of the band gap (by the LDA) affects the formation energies, as discussed in detail in the Appendix. A simple correction scheme based on the band energy has been proposed by Persson {\it et al.} \cite{PerZhaLan05} and is further motivated in the Appendix. The correction amounts to the following term which is added to the as-calculated formation energies:
\begin{align}   \Delta E^{corr}
  = (q + \Delta z_h) \Delta E_{\VB} + \Delta z_e \Delta E_{\CB},
  \label{eq:EG_corr}
\end{align}
where $\Delta E_{\VB}$ and $\Delta E_{\CB}$ are shifts of the valence and conduction band edges, respectively, to correct the band gap, while $\Delta z_h$ and $\Delta z_e$ are the number of unoccupied valence and occupied conduction band states, respectively.

\begin{figure*}
  \setlength{\myhgt}{1.1in}
  \centering
  \begin{tabular}{lcl}
    \begin{tabular}[b]{rr}
      \subfigure[\,$V_{\Al}^{3-}$]{
	\includegraphics[height=\myhgt]{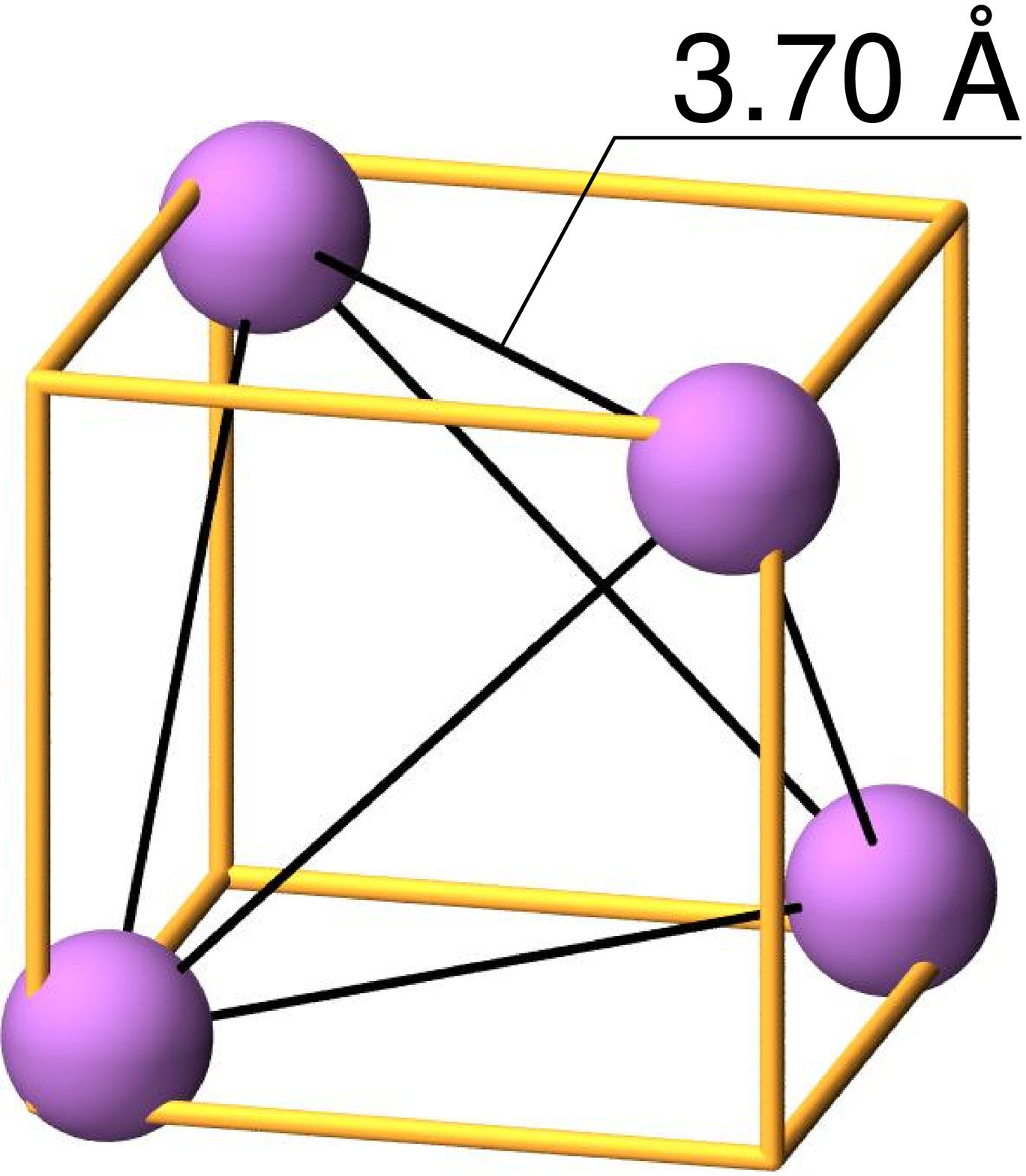}
	\label{fig:conf_VAl}}
      &
      \subfigure[\,$V_{\Sb}^{1-}$]{
	\includegraphics[height=\myhgt]{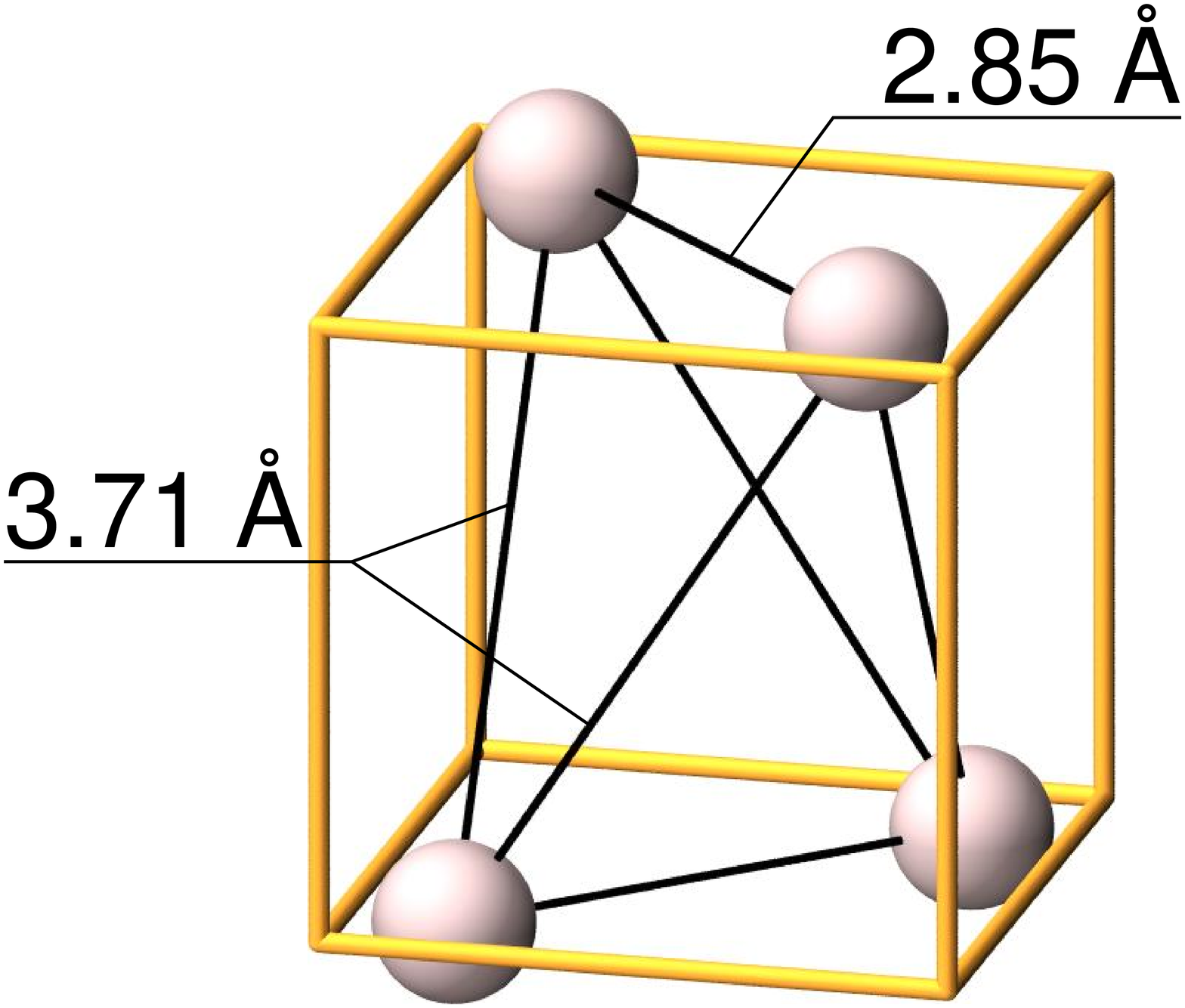}
	\label{fig:conf_VSb}}
      \\
      \subfigure[\,$\Al_{\Sb}^{2-}$]{
	\includegraphics[height=\myhgt]{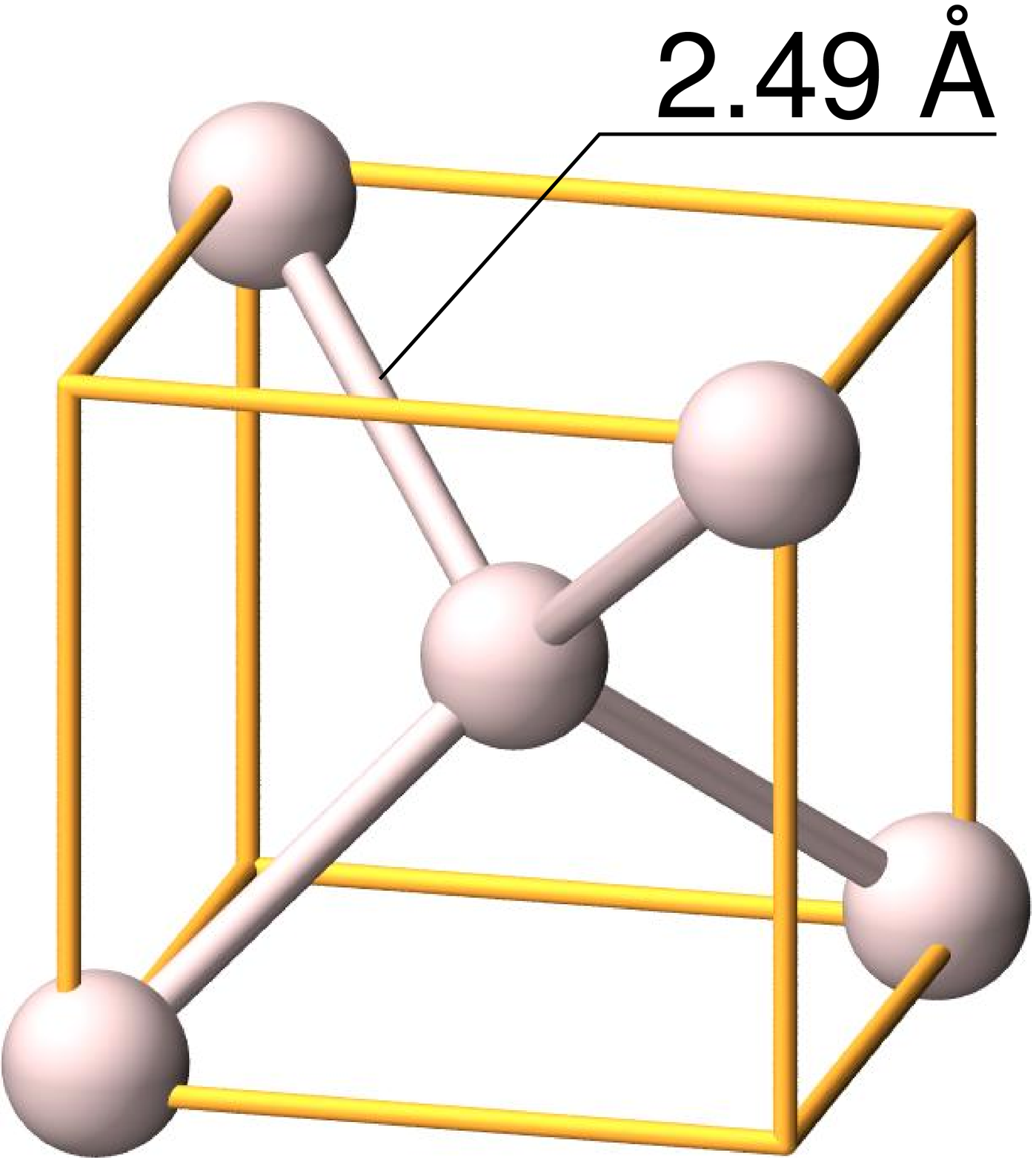}
	\label{fig:conf_AlSb}}
      &
      \subfigure[\,$\Sb_{\Al}^{0}$]{
	\includegraphics[height=\myhgt]{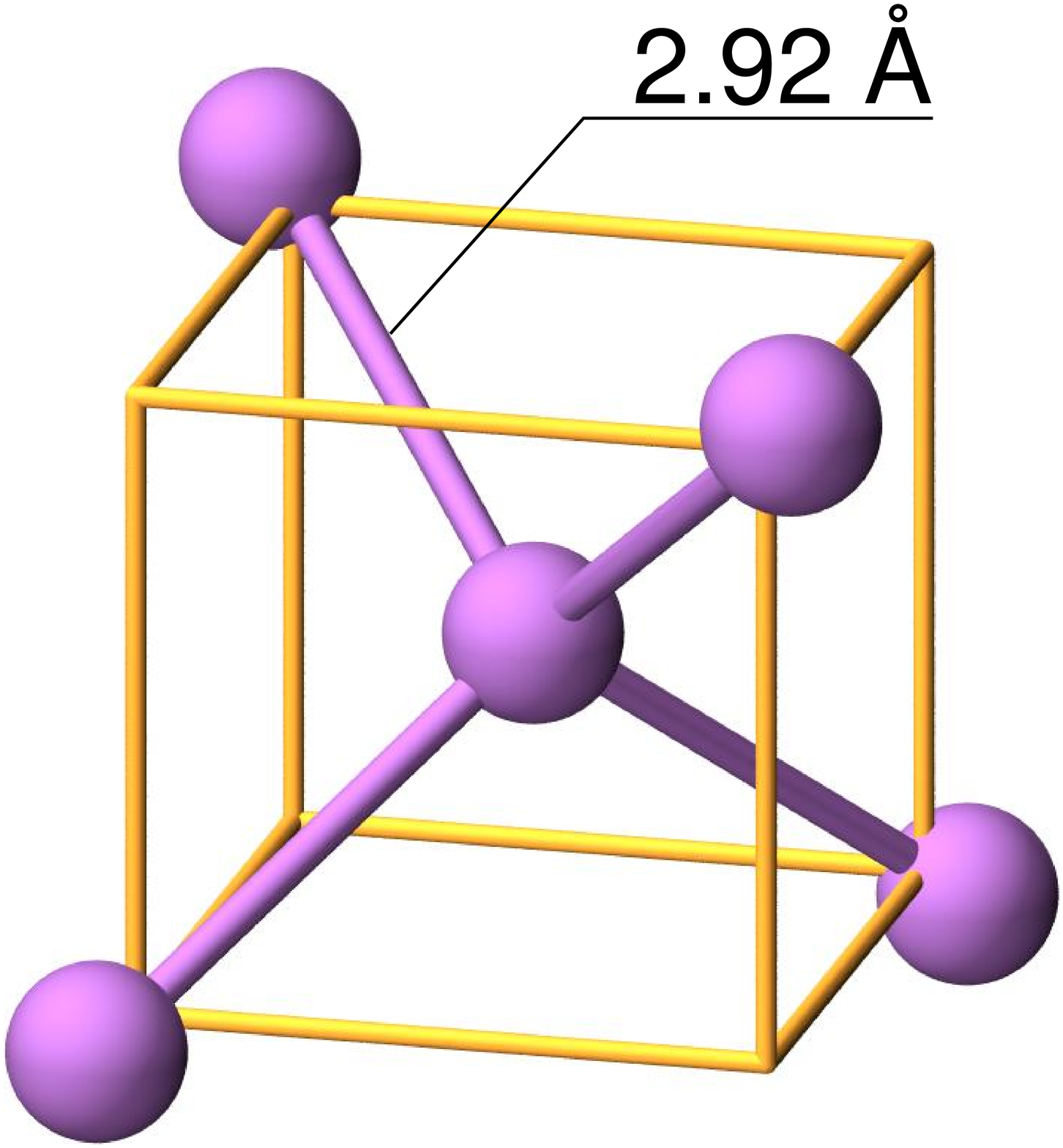}
	\label{fig:conf_SbAl}}
    \end{tabular}
    &
    \hspace{0.0in}
    \subfigure[\,conventional cell]{
      \includegraphics[width=1.7\myhgt]{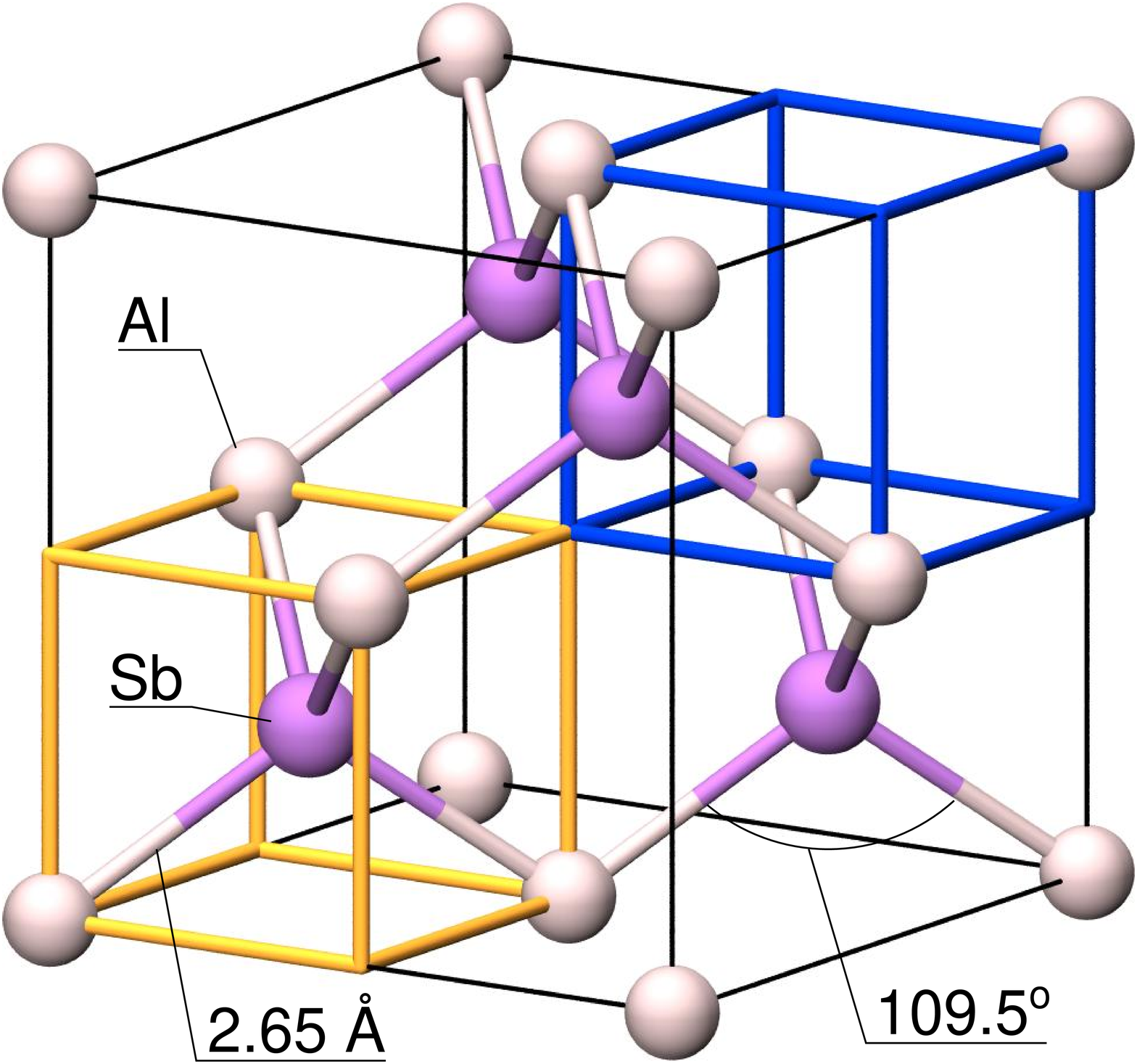}
      \label{fig:conf_unitcell}}
    \hspace{0.0in}
    &
    \begin{tabular}[b]{ll}
      \subfigure[\,$\Al_{i,\Al}^{1+}$]{
	\includegraphics[height=\myhgt]{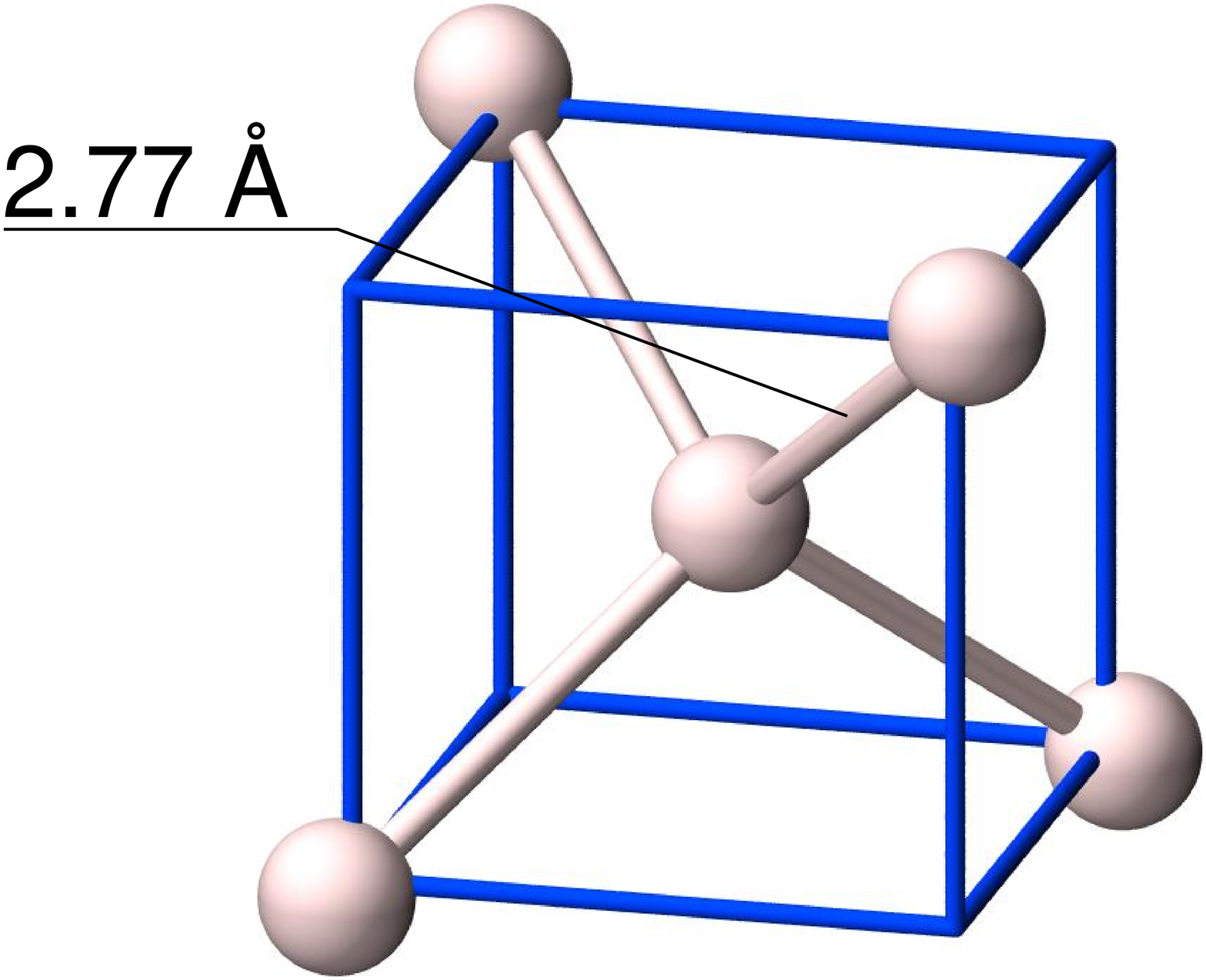}
	\label{fig:conf_Ali1}}
      &
      \subfigure[\,$\Al_{i,\Sb}^{1+}$]{
	\includegraphics[height=\myhgt]{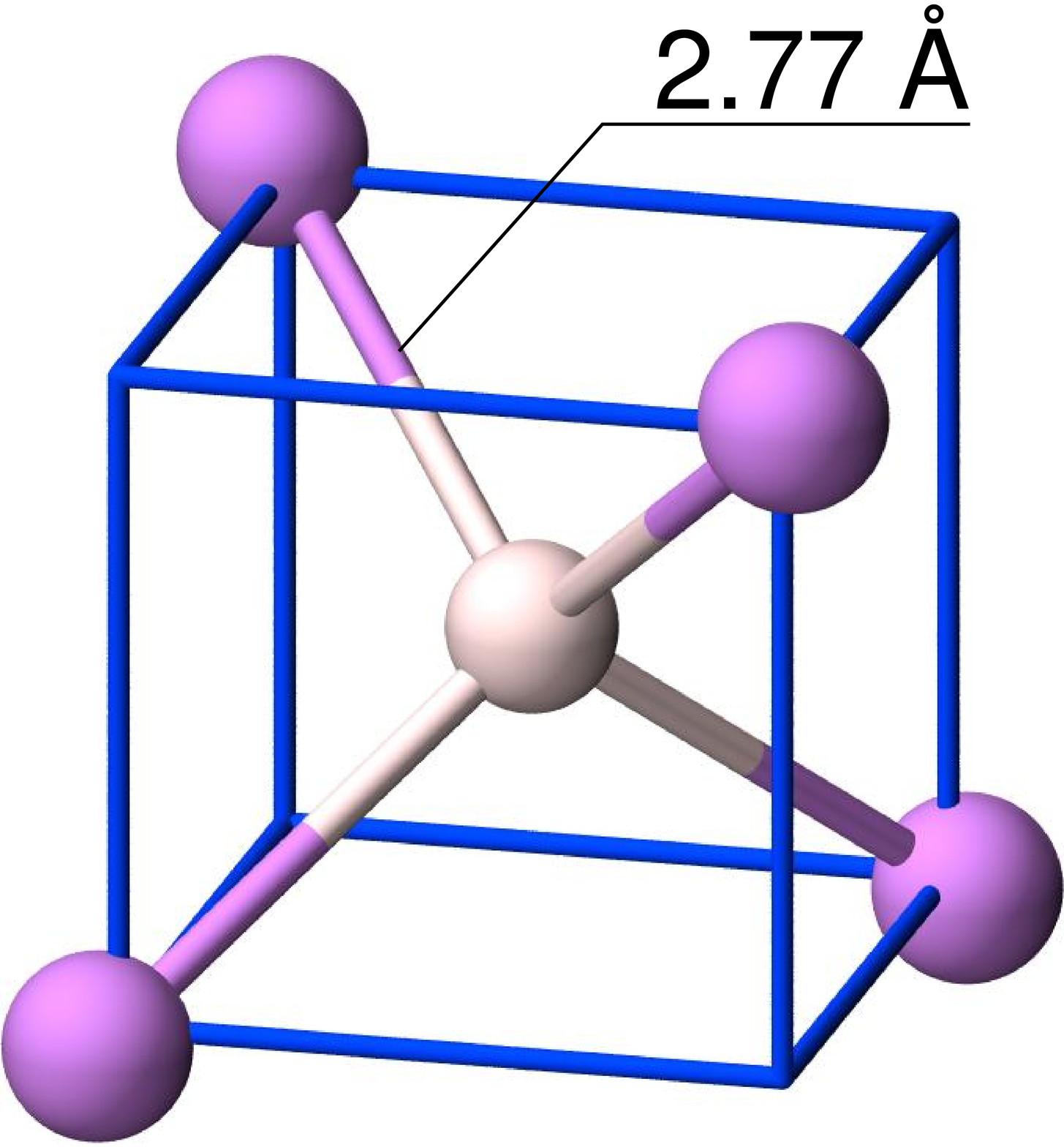}
	\label{fig:conf_Ali2}}
      \\
      \subfigure[\,$\Sb_{i,\Al}^{1+}$]{
	\includegraphics[height=\myhgt]{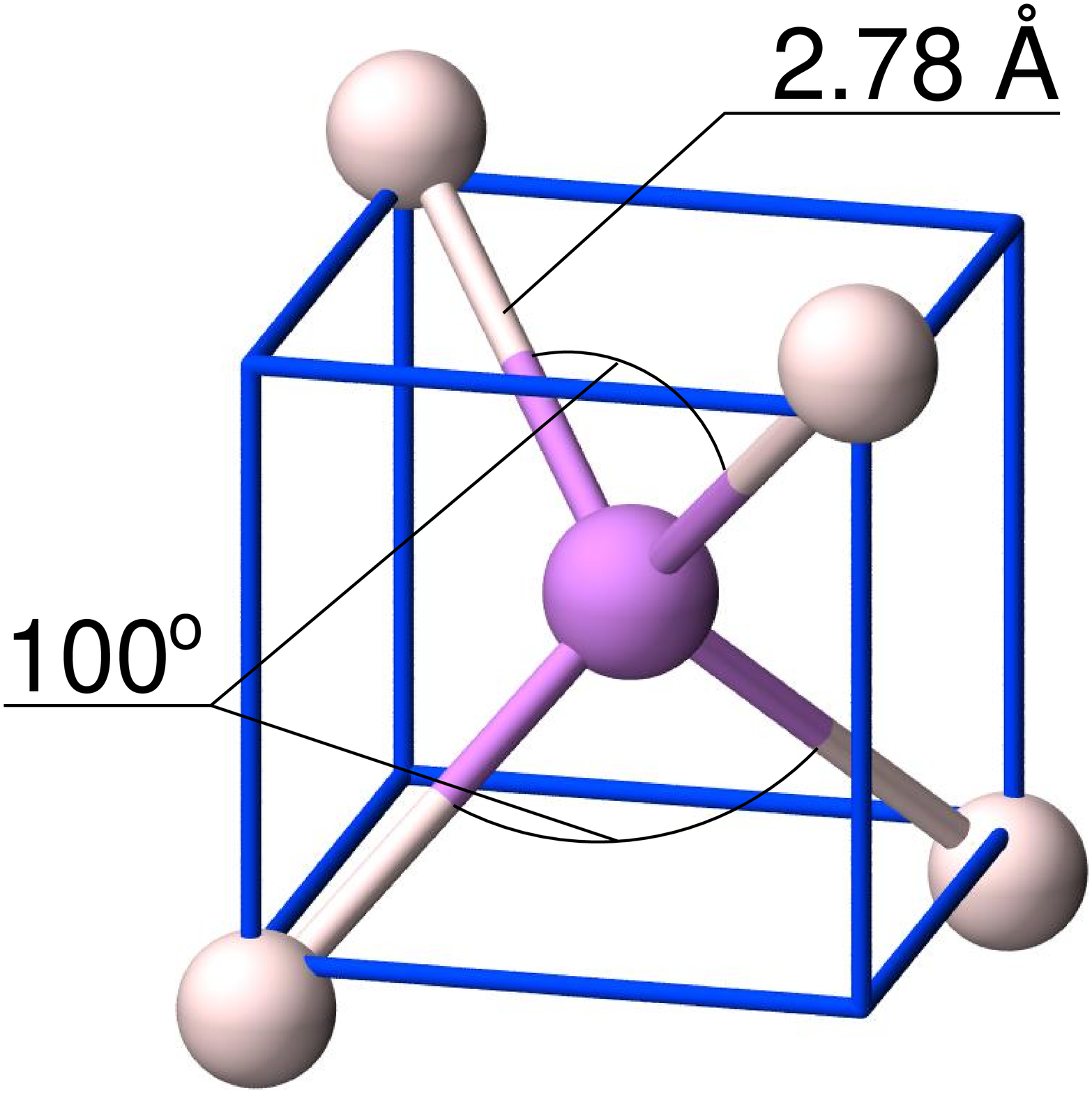}
	\label{fig:conf_Sbi2}}
      &
      \subfigure[\,$(\Al-\Al)_{\Al\left<100\right>}^0$]{
	\includegraphics[height=\myhgt]{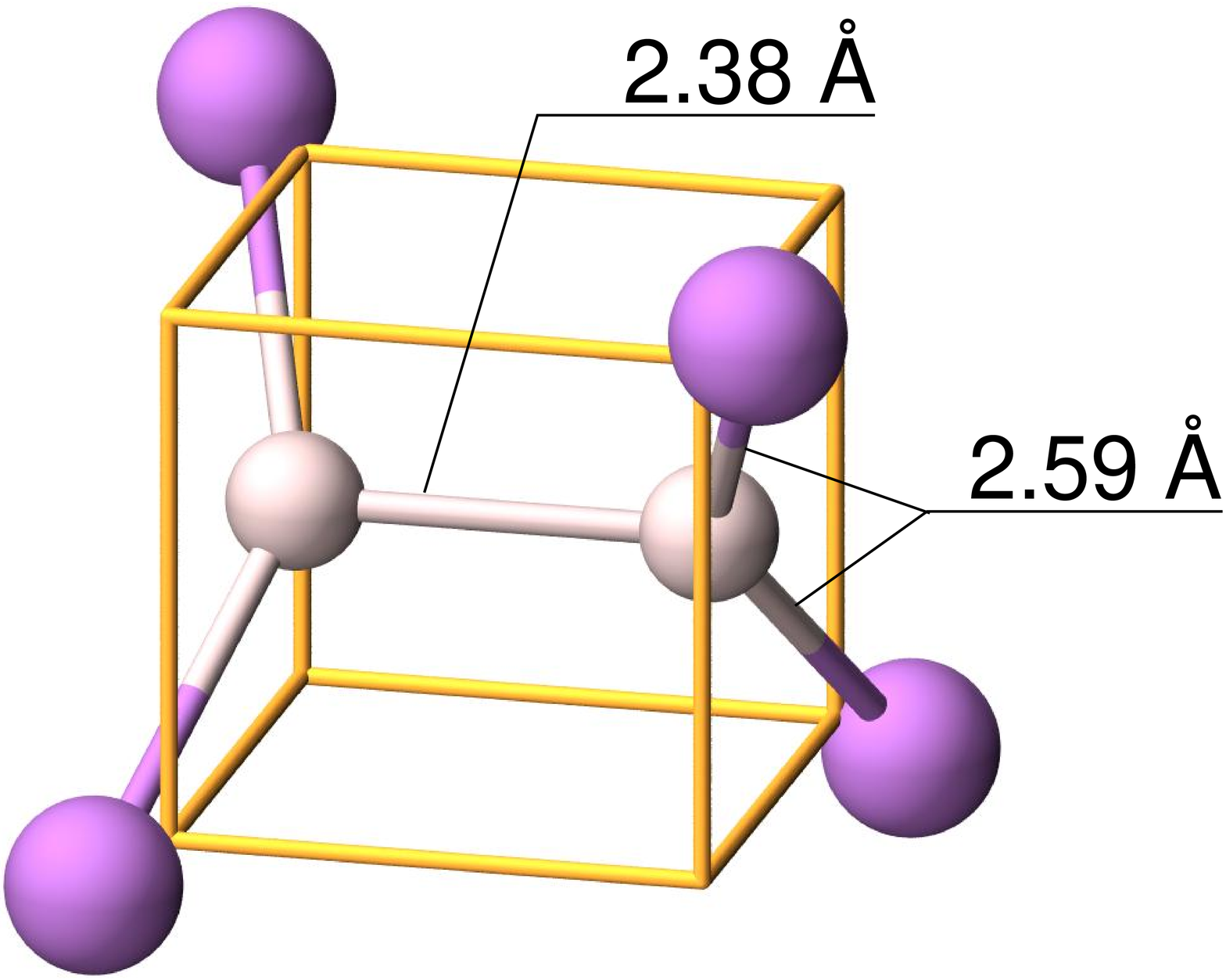}
	\label{fig:conf_AlAl_100}}
    \end{tabular}
  \end{tabular}
  \caption{
    (Color online) Relaxed defect geometries for (a,b) vacancies, (c,d) antisites, (f,g,h) several tetrahedral interstitials, and (i) one split interstitial. (e) The conventional 8-atom unit cell is shown for reference. Purple (dark grey) and beige (light grey) balls represent antimony and aluminum atoms, respectively. Note that the negatively charged antimony vacancy in (b) shows a pronounced Jahn-Teller distortion. (Indicated bond lengths were taken from 216-atom supercells.)
  }
  \label{fig:confs}

\end{figure*}

\newcommand{\stbl}{stable}
\newcommand{\fn}{$^*$}
\newcommand{\ms}{$^\dag$}
\begin{table*}
  \centering
  \caption{
    Stability of split interstitial configurations.
  }
  \label{tab:splitinterstitials}

  \begin{tabular}{cccccc}
    \hline\hline
    Charge state
    & $+2$ & $+1$ & $0$ & $-1$ & $-2$ \\
    \hline

    $(\Al-\Al)_{\Al\left<100\right>}$
    & $\Al_{i,\Sb}$\ms & \stbl & $\Al_{i,\Al}$\ms
    & $(\Al-\Al)_{\Al\left<110\right>}$\ms
    & $(\Al-\Al)_{\Al\left<110\right>}$\ms \\

    $(\Al-\Al)_{\Al\left<110\right>}$
    & $\Al_{i, \Al}$\ms & $\Al_{i,\Al}$\ms & \stbl & \stbl & \stbl \\

    $(\Al-\Sb)_{\Sb\left<100\right>}$
    & $\Al_{i,\Al}$ & $\Al_{i,\Al}$ & $\Al_{i,\Al}$
    & $\Al_{i,\Al}$ & $\Al_{i,\Al}$ \\

    $(\Al-\Sb)_{\Sb\left<110\right>}$
    & $\Al_{i,\Al}$ & \stbl & \stbl & \stbl & \stbl \\[6pt]

    $(\Sb-\Sb)_{\Sb\left<100\right>}$
    & $\Sb_{i,\Al}$\ms & $\Sb_{i,\Al}$\ms
    & $(\Sb-\Sb)_{\Sb\left<110\right>}$ & $(\Sb-\Sb)_{\Sb\left<110\right>}$
    & $(\Sb-\Sb)_{\Sb\left<110\right>}$ \\

    $(\Sb-\Sb)_{\Sb\left<110\right>}$
    & \stbl & \stbl & \stbl & \stbl & \stbl \\

    $(\Sb-\Al)_{\Al\left<100\right>}$
    & $(\Sb-\Sb)_{\Sb\left<110\right>}$ & $(\Sb-\Sb)_{\Sb\left<110\right>}$
    & \stbl \fn & \stbl \fn & \stbl \fn \\

    $(\Sb-\Al)_{\Al \left<110\right>}$
    & \stbl & \stbl & \stbl & \stbl & \stbl\\

    \hline\hline
  \end{tabular}\\
  \hspace{-10pt}\ms Relaxations starting from the idealized positions maintained the initial symmetry, but when started from\\
  \hspace{-180pt}randomized positions, relaxed to the indicated structure\\
  \hspace{-176pt}\fn Split interstitial displaced along $\left<100\right>$ from ideal position
\end{table*}

In the present work the energy offsets $\Delta E_{\VB}$ and $\Delta E_{\CB}$ were obtained from {\it G$_0$W$_0$} calculations \cite{Hed65} within the single plasmon-pole model as implemented in \textsc{abinit}. \cite{GonBeuCar02, Goe97, abinit} Non-selfconsistent {\it G$_0$W$_0$} calculations were employed to properly refer the quasiparticle energies to the same potential zero as the LDA eigenvalues.  Fritz-Haber-In\-sti\-tu\-te norm-conserving pseudopotentials \cite{FucSch99} in the Troul\-lier-Martins scheme \cite{TroMar91} were used with a cutoff energy of 15\,Hartree (Ha). The other relevant cutoff energies used in the calculation were 5\,Ha for the self-energy wave functions, 6\,Ha for the exchange part of the self-energy, and 6\,Ha for the screening matrix. The number of bands in the self-energy and screening matrix calculations were 100 and 150, respectively.

\begin{figure*}
  \setlength{\myhgt}{1.6in}
  \includegraphics[width=0.98\linewidth]{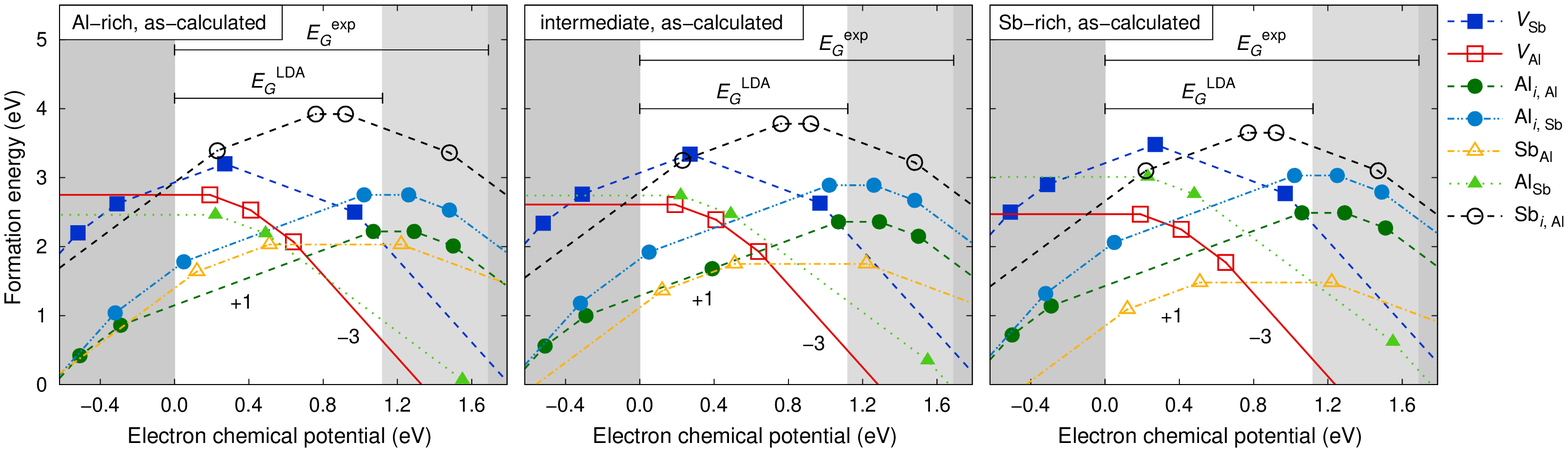} \\[10pt]
  \includegraphics[width=0.98\linewidth]{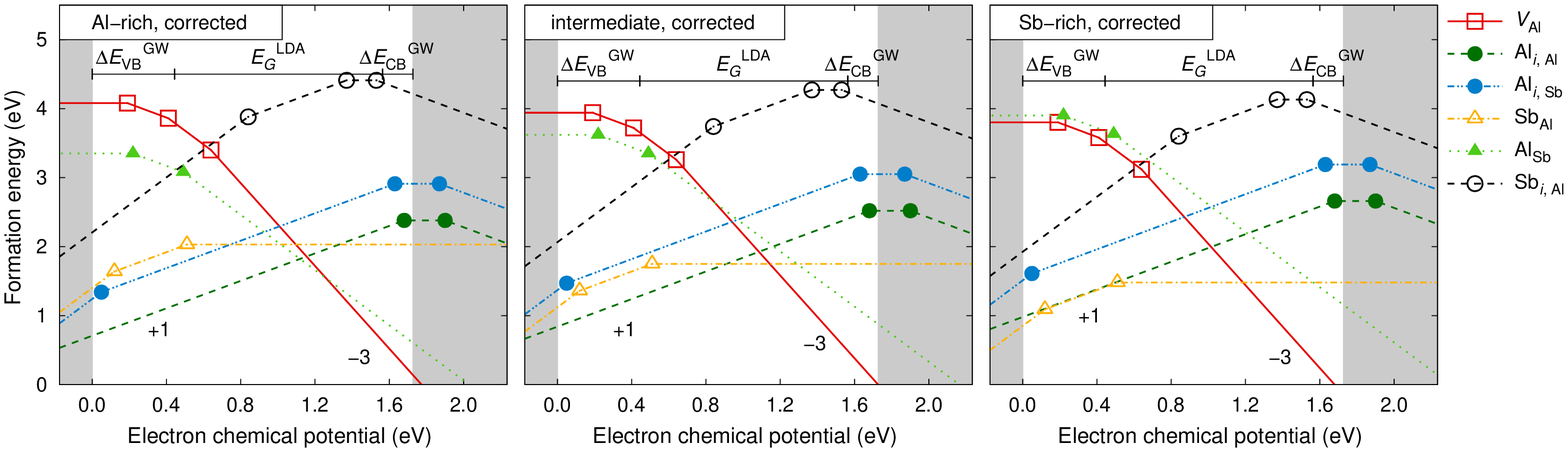}
  \caption{
    (Color online) Dependence of the formation energies on the electron chemical potential under Al-rich (left, $\Delta\mu=-\Delta H_f$), stoichiometric (center, $\Delta\mu=0\,\eV$), and Sb-rich (right, $\Delta\mu=+\Delta H_f$) conditions for the most important point defects. The top panels show the formation energies based on the as-calculated values, while the bottom panels show the results of applying the band gap correction term of \eq{eq:EG_corr}. The slope of each line is determined by the charge state, according to \eq{eq:eform}. ($\Delta H_f^\text{calc}=-0.28\,\eV$)
  }
  \label{fig:eform}
\end{figure*}

In \tab{tab:eform1} we report both the as-calculated formation energies (including finite-size scaling) and the band gap-corrected formation energies. The correction terms can be reproduced using the values for $\Delta z_h$ and $\Delta z_e$ included in the table. Note that we extracted $\Delta z_h$ and $\Delta z_e$ values only for the most important defects for which an unambiguous distinction between valence and conduction band states was possible. A direct comparison of the as-calculated and band gap-corrected values is shown in \fig{fig:eform}, and discussed later in \sect{sec:eform}.

The defect concentrations calculated in \sect{sect:conc} were obtained using the band gap-corrected formation energies and the {\it GW} band gap.

\section{Results}
\label{sect:results}

\subsection{Bulk properties}

As described in \sect{sect:method_eform} and evident in \eq{eq:eform}, the determination of defect formation energies requires that we also calculate bulk properties of the solid and its constituents in reference states. The reference states for Al and Sb are face-centered cubic (fcc) and rhombohedral solids, respectively.

For fcc aluminum, we obtain a lattice constant of $3.99\,\AAA$ and a cohesive energy of $-4.19\,\eV/\atom$ which compare reasonably well with the experimental values of $4.05\,\AAA$ (room temperature)\cite{LattConsts} and $-3.38\,\eV/\atom$.\cite{CohesiveEnergies} The underestimation of the lattice constant and the overestimation of the cohesive energy are typical for LDA calculations. Antimony has a rhombohedral groundstate structure (R$\bar{3}$mh, space group no.~166, {\it Strukturbericht} symbol A7) for which the DFT calculations yield a lattice constant of $4.46\,\AAA$ and a rhombohedral angle of $59.0\deg$ (experimental values: $4.50\,\AAA$ and $57.1\deg$, Ref.~[\onlinecite{LattConsts}]) and a cohesive energy of $-4.81\,\eV/\atom$ (experimental value: $-2.72\,\eV/\atom$, Ref.~[\onlinecite{CohesiveEnergies}]).

At ambient conditions bulk aluminum antimonide adopts the zinc-blende structure (F$\bar{4}$3m, space group no.~216, {\it Strukturbericht} symbol B3). The DFT calculated lattice constant is 6.12\,\AA\ in good agreement with the experimental value of 6.13\,\AA\ (300\,K). The DFT calculations furthermore yield a formation enthalpy of $-0.28\,\eV/\text{f.u.}$ (experimental value: $-0.84\,\eV/\text{f.u.}$). The direct and indirect band gaps are calculated as 1.53 and 1.12\,eV, respectively, while the experimental values are 2.30 and 1.62\,eV at 300\,K.\cite{MadelungSemiHandbook}

From the {\it GW} calculations, we obtain direct and indirect band gaps of 2.31 and 1.64\,eV, respectively, in excellent agreement with the experimental values cited above. The shifts of the valence and conduction band edges showed small variations with $k$-vector, so we calculated $\Delta E_{\VB}$ and $\Delta E_{\CB}$ which appear in \eq{eq:EG_corr} as weighted averages over all $k$-points included in the calculations. The shifts thus obtained are $\Delta E_{\VB}=-0.44\,\eV$ and $\Delta E_{\CB}=0.16\,\eV$.

\subsection{Defect structures}

We considered all possible native defects in AlSb up to split interstitials. In total, this amounts to 18 different defects (two vacancies, two antisite defects, four tetrahedral interstitials, two hexagonal interstitials, and eight split interstitials), not all of which are stable. For each defect, we investigated a series of charge states, generally from $q = -3$ to $q= +3$, as appropriate. Defect complexes were not considered.

\textit{Vacancies.}
In the ideal zinc-blende structure, both Al and Sb sites possess tetrahedral symmetry with nearest neighbor distance of 2.65\,\AA. If one removes a single atom and allows the system to relax from randomized positions, the aluminum vacancy $V_{\Al}$ maintains the $T_d$ symmetry for all relevant charge states and the surrounding antimony ions relax inward by 0.36\,\AA\ (for $q=0$) to 0.38\,\AA\ (for $q=-3$) [see for example \fig{fig:conf_VAl}]. This behavior is typical of cation vacancies in III-V and II-VI zinc-blende semiconductors, where a dimerization transformation of the atoms surrounding the vacancy is generally not energetically favorable.\cite{ChadiVacancyStruct} In contrast, the antimony vacancy $V_{\Sb}$ exhibits a Jahn-Teller distortion to a local tetragonal symmetry for all but the positive charge states. The relaxed configuration for $V_{\Sb}^{3+}$ is shown in \fig{fig:conf_VSb} representatively, also indicating the pairing of Al atoms in the first neighbor shell of the vacancy.

\textit{Antisites.}
Both the $\Al_{\Sb}$ and $\Sb_{\Al}$ antisite defects maintain the $T_d$ symmetry for all relevant charge states. In the case of $\Al_{\Sb}$, the surrounding aluminum ions relax inwards, while for $\Sb_{\Al}$ the nearest neighbor antimony ions relax outwards [compare Figures~\ref{fig:confs}(c) and (d)]. These relaxations occur as expected based on the atomic radii.

\textit{Tetrahedral interstitials.}
In the zinc-blende structure there are two distinct tetrahedral sites: one centered on an Al tetrahedron (4d site) and one centered on an Sb tetrahedron (4b site). Thus, there are four possible types of tetrahedral interstitials: $\Al_{i,\Al}$ and $\Sb_{i,\Al}$ on the 4d site; $\Al_{i,\Sb}$ and $\Sb_{i,\Sb}$ on the 4b site.

Both kinds of tetrahedral aluminum interstitials ($\Al_{i,\Al}$, $\Al_{i,\Sb}$) maintain the $T_d$ symmetry after relaxation, with the neighboring ions relaxing outwards as shown in Figs.~\ref{fig:confs}(f) and (g). The $\Sb_{i,\Sb}$ configuration is unstable in all but the $+2$ charge state; in all other charge states, the interstitial atom relaxes either onto a hexagonal site or forms a split interstitial. Conversely, $\Sb_{i,\Al}$ is stable in all of its charge states, but does exhibit Jahn-Teller distortions as illustrated in \fig{fig:conf_Sbi2}.

\textit{Hexagonal interstitials.}
The hexagonal interstitial is located at Wyckoff position 16e. The aluminum hexagonal interstitial $\Al_{i,\text{hex}}$ was found to be unstable for all charge states, with the Al atom observed to relax into the $\Al_{i,\Al}$ configuration even when starting from ideal positions. For the antimony hexagonal interstitial $\Sb_{i,\text{hex}}$, the $+2$ charge state was found to relax directly into the $\Sb_{i,\Al}$ position, whereas for charge states $+1$ and $-2$ the interstitial atom remained in the hexagonal site. The other charge states when perturbed from the ideal position relax into the $(\Sb-\Sb)_{\Sb\langle 110\rangle}$ split interstitial, described below.

\textit{Split interstitials.}
Split interstitial configurations, in which two atoms share one atom site, have been extensively discussed in the literature for the zinc-blende structure. \cite{BarSch85, ZhaNor91, Cha92, LaaNiePus92, SchMorPap02, ZollNie03, ElMMou05} We have carried out an exhaustive exploration of the structures of these defects in AlSb. The following cases were considered: two Al atoms oriented along $\left<100\right>$ sharing one Al site [$(\Al-\Al)_{\Al\left<100\right>}$], the same atoms but oriented along $\left<110\right>$ [$(\Al-\Al)_{\Al\left<110\right>}$], one Al atom and one Sb atom oriented along $\left<100\right>$ sharing one Al site [$(\Sb-\Al)_{\Al\left<100\right>}$], and the same combination of atoms oriented along $\left<110\right>$ [$(\Sb-\Al)_{\Al\left<110\right>}$]. Four more configurations are obtained corresponding to the same combinations above, but sharing an Sb site.

Most of these configurations actually are found to be unstable with respect to other interstitial configurations. The results of these calculations are summarized in \tab{tab:splitinterstitials}, which shows which configurations were found to be stable and which ones were unstable or only conditionally stable with respect to alternative interstitial configurations. In certain cases, a structure relaxed starting from the ideal atomic coordinates maintained the starting symmetry after relaxation, but when started from randomized coordinates, the structure relaxed to a different configuration; these cases are indicated in \tab{tab:splitinterstitials} by listing both final configurations. The cases simply marked stable (single entry `\stbl') in \tab{tab:splitinterstitials} relaxed to the ideal symmetry configuration for all starting configurations.

For completeness, we note that the split interstitial $(\Sb\!\!~-~\!\!\Al)_{\Al\left<100\right>}$ in charge states $q= 0$, $-1$, and $-2$ is displaced along $\left<100\right>$ from the ideal position, such that the Sb interstitial is located in between two regular Sb atoms along the $\left<110\right>$ direction.

\subsection{Formation energies}
\label{sec:eform}

\begin{table}
  \caption{
    Formation energies, in eV, of intrinsic point defects under Al-rich ($\Delta\mu=-\Delta H_f$) and Sb-rich ($\Delta\mu=+\Delta H_f$) conditions for an electron chemical potential at the valence band maximum [$\mu_e=0\,\eV$ in equation \eq{eq:eform}]. Both the as-calculated and the band-gap corrected formation energies are given. The number of occupied (unoccupied) conduction (valence) band states defined through \eq{eq:z_eh} is given in the third column ($\Delta z_{e,h}$), where negative (positive) values correspond to electrons (holes). The extrapolation error, in eV, of the finite-size scaling procedure is given in brackets in the last column.
  }
  \label{tab:eform1}
  \centering
  \newcolumntype{Q}[0]{>{(}d<{)}}
  \newcommand{\spread}[1]{\multicolumn{1}{c}{#1}}
  \newcommand{\spreadd}[1]{\multicolumn{2}{c}{#1}}
  \begin{tabular}{lddddddQ}
    \hline\hline
    \multicolumn{1}{l}
    {Defect}
    & \spread{$q$}
    & \spread{$\Delta z_{e,h}$}
    & \spreadd{as-calc.}
    & \spreadd{corr.}
    & \spread{}
    \\
    &
    &
    & \spread{Al-rich}
    & \spread{Sb-rich}
    & \spread{Al-rich}
    & \spread{Sb-rich}
    & \spread{}
    \\

    \hline
    
    $V_{\Sb}$
    & -3 &   - &   5.42 &   5.70 &  -    & -    & 0.04 \\
    & -2 &   - &   4.48 &   4.76 &  -    & -    & 0.04 \\
    & -1 &   0 &   3.47 &   3.75 &  3.91 & 4.19 & 0.08 \\
    & 0  &   - &   3.29 &   3.56 &  -    & -    & 0.07 \\
    & +1 &   0 &   2.93 &   3.21 &  2.49 & 2.76 & 0.04 \\
    & +2 &   - &   3.24 &   3.52 &  -    & -    & 0.08 \\
    & +3 &   - &   3.76 &   4.03 &  -    & -    & 0.12 \\[6pt]
    
    $\Al_{\Sb}$
    & -3 &    - &   4.72 &   5.27 &  -    & -    & 0.05 \\
    & -2 &    0 &   3.17 &   3.72 &  4.06 & 4.61 & 0.02 \\
    & -1 &   +1 &   2.68 &   3.24 &  3.57 & 4.12 & 0.01 \\
    & 0  &   +2 &   2.46 &   3.01 &  3.35 & 3.90 & 0.03 \\[6pt]
    
    $\Al_{i,\Al}$
    & -2 &    - &   5.01 &   5.29 &  -    & -    & 0.06 \\
    & -1 &   -2 &   3.51 &   3.78 &  4.28 & 4.56 & 0.03 \\
    & 0  &   -1 &   2.22 &   2.49 &  2.38 & 2.66 & 0.01 \\
    & +1 &    0 &   1.15 &   1.43 &  0.71 & 0.98 & 0.01 \\
    & +2 &   +1 &   1.44 &   1.72 &  1.00 & 1.27 & 0.02 \\
    & +3 &    - &   1.95 &   2.22 &  -    & -    & 0.06 \\[6pt]
    
    $\Al_{i,\Sb}$
    & -2 &    - &   5.49 &   5.77 &  -    & -    & 0.07 \\
    & -1 &   -2 &   4.01 &   4.28 &  4.78 & 5.06 & 0.03 \\
    & 0  &   -1 &   2.75 &   3.03 &  2.91 & 3.19 & 0.01 \\
    & +1 &    0 &   1.73 &   2.01 &  1.29 & 1.56 & 0.01 \\
    & +2 &   +1 &   1.68 &   1.96 &  1.24 & 1.51 & 0.01 \\
    & +3 &    - &   2.00 &   2.28 &  -    & -    & 0.01 \\[6pt]

    $V_{\Al}$
    & -3 &    0 &   3.99 &   3.72 &  5.32 & 5.04 & 0.02 \\
    & -2 &   +1 &   3.35 &   3.07 &  4.68 & 4.40 & 0.02 \\
    & -1 &   +2 &   2.94 &   2.66 &  4.27 & 3.99 & 0.03 \\
    & 0  &   +3 &   2.75 &   2.47 &  4.08 & 3.80 & 0.04 \\[6pt]
    
    $\Sb_{\Al}$
    & -1 &    - &   3.25 &   2.70 &  -    & -    & 0.03 \\
    & 0  &    0 &   2.03 &   1.48 &  2.03 & 1.48 & 0.02 \\
    & +1 &   +1 &   1.52 &   0.97 &  1.52 & 0.97 & 0.01 \\
    & +2 &   +2 &   1.40 &   0.85 &  1.40 & 0.85 & 0.01 \\[6pt]
    
    $\Sb_{i,\Sb}$
    & +2 &    - &   3.63 &   3.35 &  -    & -    & 0.01 \\[6pt]
    
    $\Sb_{i,\Al}$
    & -2 &    - &   6.32 &   6.04 &  -    & -    &  0.10 \\
    & -1 &   -4 &   4.84 &   4.57 &  5.94 & 5.66 &  0.02 \\
    & 0  &   -3 &   3.92 &   3.65 &  4.41 & 4.13 & <0.01 \\
    & +1 &   -2 &   3.16 &   2.88 &  3.04 & 2.77 &  0.13 \\
    & +2 &   -1 &   2.93 &   2.66 &  2.21 & 1.93 &  0.01 \\[6pt]
    
    $\Sb_{i,hex}$
    & -2 &    - &   6.24 &   5.96 &  -    & -    & 0.09 \\
    & -1 &   -4 &   4.96 &   4.68 &  6.06 & 5.78 & 0.02 \\
    & 0  &   -3 &   3.93 &   3.66 &  4.42 & 4.14 & 0.01 \\
    & +1 &   -2 &   3.14 &   2.86 &  3.02 & 2.75 & 0.01 \\

    \hline\hline

  \end{tabular}

\end{table}

\begin{table}

  \caption{
    As-calculated formation energies for the split-interstitial configurations which were found to be stable or metastable. Values are given for both Al-rich and Sb-rich conditions. The extrapolation error of the finite-size scaling procedure is given in brackets in the last column. Values are in eV.
  }

  \label{tab:eform2}

  \newcolumntype{Q}[0]{>{(}d<{)}}
  \newcommand{\spread}[1]{\multicolumn{1}{c}{#1}}
  \newcommand{\spreadd}[1]{\multicolumn{2}{c}{#1}}

  \begin{tabularx}{0.9\columnwidth}{XdddQ}
    \hline\hline

    \multicolumn{1}{l}{Defect}
    & \spread{$q$}
    & \spread{Al-rich}
    & \spread{Sb-rich}
    & \spread{}
    \\
    
    \hline
    
    $(\Al-\Al)_{\Al\left<100\right>}$
    & -2 &   6.93 &   7.20 &  0.11 \\
    & -1 &   5.48 &   5.76 &  0.08 \\
    & 0  &   4.27 &   4.54 &  0.06 \\
    & +1 &   3.31 &   3.58 &  0.06 \\
    & +2 &   3.38 &   3.65 &  0.02 \\[6pt]
    
    $(\Al-\Al)_{\Al\left<110\right>}$
    & -2 &   4.74 &   5.02 &  0.04 \\
    & -1 &   3.32 &   3.60 &  0.02 \\
    & 0  &   2.68 &   2.96 & <0.01 \\
    & +1 &   2.36 &   2.64 &  0.01 \\
    & +2 &   2.54 &   2.82 &  0.01 \\[6pt]
    
    $(\Al-\Sb)_{\Sb\left<110\right>}$
    & -2 &   5.68 &   5.96 &  0.06 \\
    & -1 &   4.34 &   4.61 &  0.04 \\
    & 0  &   3.34 &   3.62 &  0.01 \\
    & +1 &   2.68 &   2.96 &  0.01 \\
    & +2 &   1.68 &   1.96 &  0.01 \\[6pt]
    
    $(\Sb-\Al)_{\Al\left<100\right>}$
    & -2 &   5.94 &   5.67 &   0.11 \\
    & -1 &   4.53 &   4.26 &   0.09 \\
    & 0  &   3.91 &   3.63 &   0.03 \\[6pt]
    
    $(\Sb-\Al)_{\Al\left<110\right>}$
    & -2 &   5.70 &   5.42 &  0.08 \\
    & -1 &   4.26 &   3.98 &  0.05 \\
    & 0  &   3.46 &   3.19 &  0.01 \\
    & +1 &   3.01 &   2.74 &  0.07 \\
    & +2 &   3.00 &   2.72 &  0.01 \\[6pt]
    
    $(\Sb-\Sb)_{\Sb\left<100\right>}$
    & +1 &   3.72 &   3.44 &  0.02 \\
    & +2 &   3.72 &   3.44 &  0.01 \\[6pt]
    
    $(\Sb-\Sb)_{\Sb\left<110\right>}$
    & -2 &   5.13 &   4.85 &  0.11 \\
    & -1 &   3.72 &   3.45 &  0.08 \\
    & 0  &   3.28 &   3.00 &  0.05 \\
    & +1 &   3.17 &   2.90 &  0.03 \\
    & +2 &   3.45 &   3.18 &  0.04 \\

    \hline\hline

  \end{tabularx}

\end{table}

The defect formation energies calculated using \eq{eq:eform} for Al-rich and Sb-rich conditions and an electron chemical potential at the valence band maximum are given in Tables~\ref{tab:eform1} and \ref{tab:eform2}. The formation energies for the dominant defects (lowest $\Delta E_D$) are shown as a function of the electron chemical potential in \fig{fig:eform}.

Figure~\ref{fig:eform} also illustrates the effect of the band gap correction described in \sect{sect:EG_corr}. In the top row we show the formation energies after the finite-size scaling procedure was applied but without the band gap correction. The bottom row shows the results including the band gap corrections. It is apparent that as the band gap correction is applied donor and acceptor levels track the conduction and valence band edges, respectively. This feature is independent of the relative values of $\Delta E_{\VB}$ and $\Delta E_{\CB}$ in \eq{eq:EG_corr}.

It is interesting to compare the band gap-corrected formation energies in the bottom panel of \fig{fig:eform} with those in the top panel if the band gap is simply extended to the experimental value by shifting the conduction band upwards. Considerable differences are observed between the two cases, with the values corrected using the scheme presented here being more consistent. In fact, we observe that, qualitatively, the results obtained by applying no correction at all are more similar to the corrected results than what is obtained by assigning the band gap error fully to the conduction band.\cite{FosSulLop01, KohCedMor00} Furthermore, a correction procedure often suggested in the literature\cite{SchMorPap02} to simply shift the ionization energies of donor-like defects to track the conduction band minimum (implicitly leaving the ionization energies of acceptor-like defects tracking the valence band maximum), without accounting for the occupation of states, does not properly correct the errors in the formation energies. The scheme employed here results in an identical shift of the ionization energies, but also corrects the errors in the formation energies.However, we note that this scheme still neglects both level relaxations and changes in the double counting term.

We believe the method presented here is the most consistent way to address the LDA band gap problem, with the use of {\it GW} calculations providing a first principles approach to calculating the correction terms. We find that neglecting to include the band gap correction terms in the formation energies leads to significant errors in the prediction of defect concentrations and which defects are dominant.

\subsection{Defect and charge carrier concentrations}
\label{sect:conc}

With the formation energies known, the equilibrium defect concentrations for a given chemical potential difference $\Delta \mu$ can be calculated using \eq{eq:concentration}. The defect concentrations depend on the electron chemical potential via \eq{eq:eform}. In the absence of extrinsic defects, the electron chemical potential is constrained by the charge neutrality condition
\begin{align}
  0 = n_e - n_h - \sum_i^{\text{defects}} q_i c_i, 
  \label{eq:chgneutral}
\end{align}
since the intrinsic concentrations of electrons and holes, $n_e$ and $n_h$, respectively, are given by
\begin{subequations}
  \begin{align}
    n_e &= \int D(E) f(E;\mu_e) dE\\
    n_h &= \int D(E) \left[1-f(E;\mu_e)\right] dE.
  \end{align}
  \label{eq:np}
\end{subequations}
Here, $D(E)$ is the electronic density of states and $f(E;\mu_e) = \left\{ 1 + e^{\left(E-\mu_e\right)/k_B T} \right\}^{-1}$ is the Fermi-Dirac distribution. The implicit dependence of the charge neutrality condition on the electron chemical potential $\mu_e$ is apparent from Eqs.~\eqref{eq:np}. To obtain the charge carrier and defect concentrations, then, we must iteratively solve \eq{eq:chgneutral} to self-consistently determine the intrinsic electron chemical potential.

\begin{figure}
  \includegraphics[width=0.98\columnwidth]{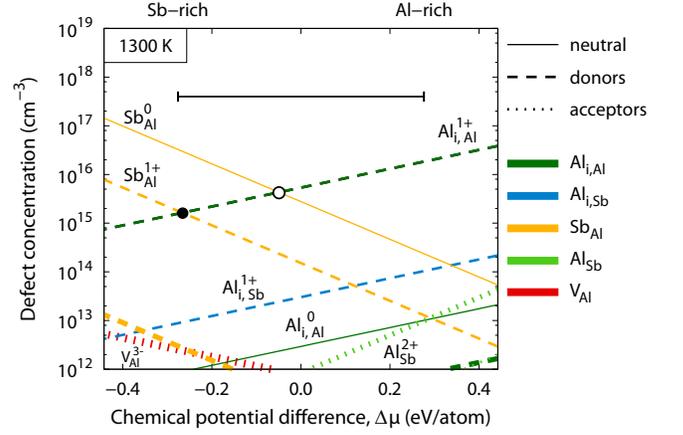}
  \caption{
    (Color online) Concentrations of individual defects at a temperature of 1300\,K as a function of the chemical potential difference $\Delta \mu$. Line thicknesses scale with charge state. Under Sb-rich conditions the dominant defect is the neutral $\Sb_{\Al}^0$ antisite, whereas under Al-rich conditions the positively charged aluminum interstitial $\Al_{i,\Al}^{1+}$ is dominant. The crossover point between $\Al_{i,\Al}^{1+}$ and $\Sb_\Al^0$ (open circle) determines the minimum in the total defect concentration shown in \fig{fig:defectsdmu}, while the crossover point between $\Al_{i,\Al}^{1+}$ and $\Sb_\Al^{1+}$ (filled circle) determines the minimum in the net electron concentration shown in \fig{fig:dconcdmu}. The horizontal bar marks the calculated range of variation of $\Delta \mu$ given by the formation enthalpy of the compound ($-\Delta H_f \leq \Delta\mu \leq +\Delta H_f$, with $\Delta H_f^\text{calc}=-0.28\,\eV$).
  }
  \label{fig:defectstypes}
\end{figure}

\begin{figure*}
  \subfigure{
    \includegraphics[width=0.9\columnwidth]{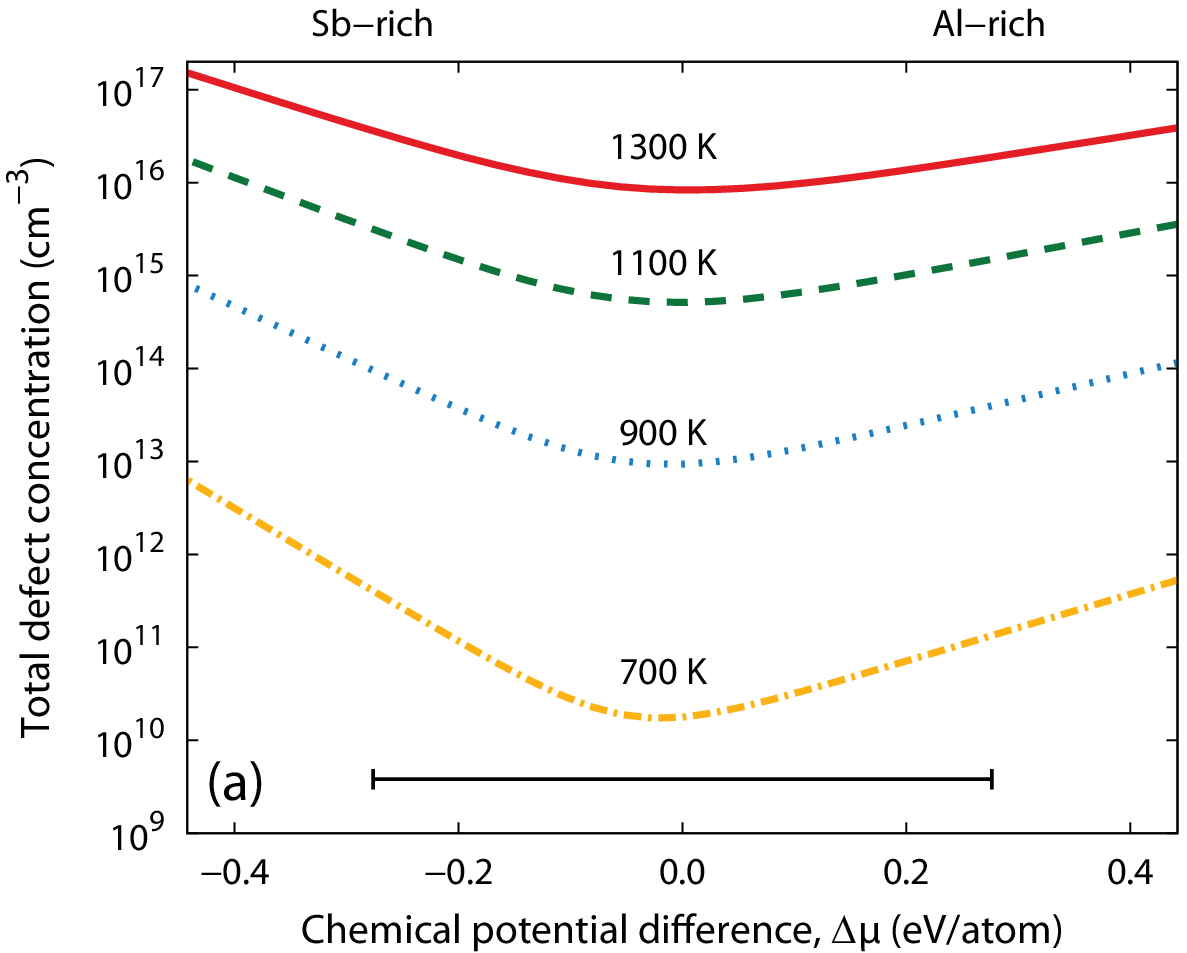}
    \label{fig:defectsdmu}
  }
  \subfigure{
    \includegraphics[width=0.9\columnwidth]{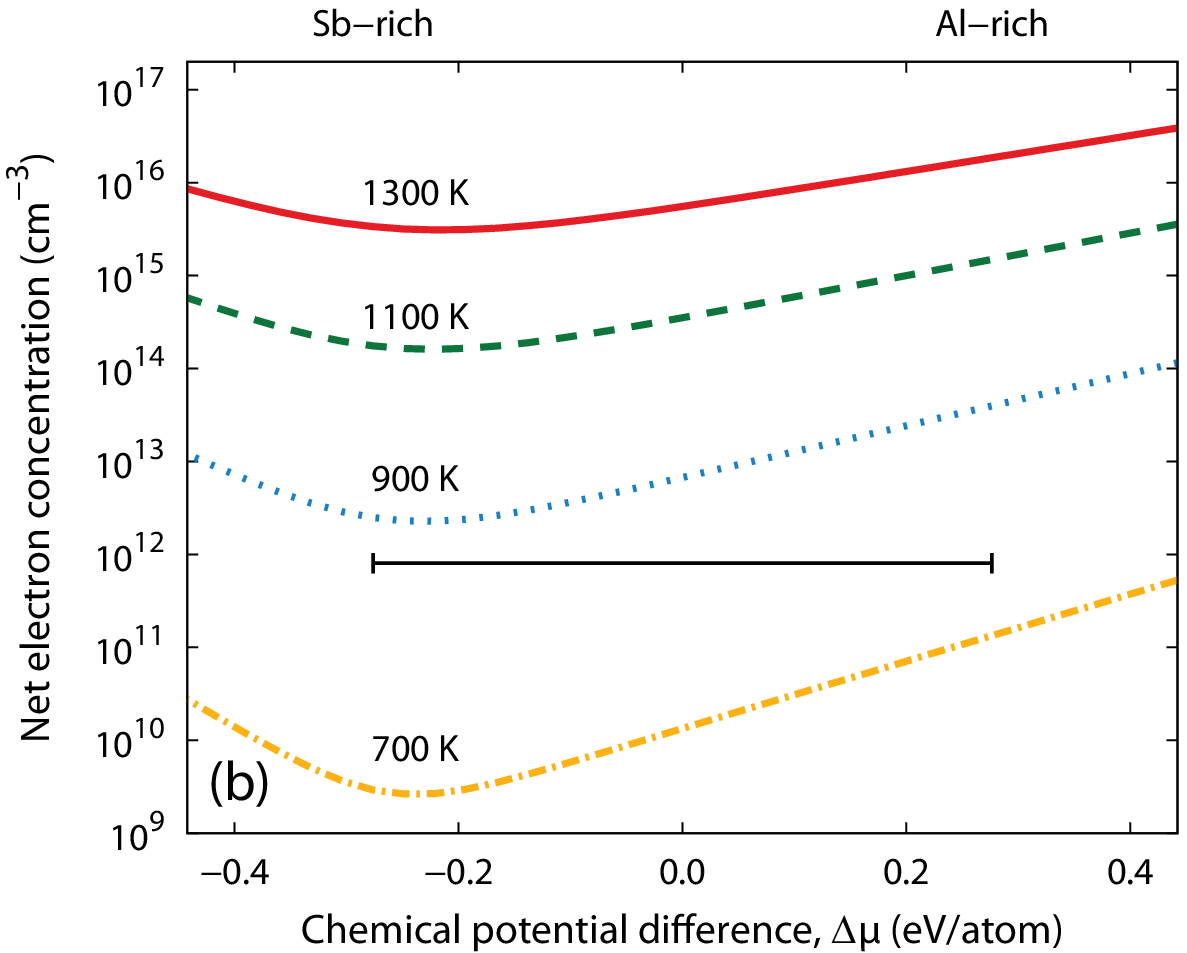}
    \label{fig:dconcdmu}
  }
  \caption{
    (Color online) (a) Equilibrium total defect concentration and (b) net electron concentration ($n_e-n_h$) at different temperatures as a function of the chemical potential difference. Note that the minimum in the total defect concentration corresponds to the crossover point between $\Al_{i,\Al}^{1+}$ and $\Sb_{\Al}^{0}$, while the minimum in the net electron concentration corresponds to the crossover point between $\Al_{i,\Al}^{1+}$ and $\Sb_{\Al}^{1+}$, as indicated in \fig{fig:defectstypes}. The dominant defects on the Sb-rich side are $\Sb_{\Al}$ and on the Al-rich side are $\Al_{i,\Al}$. The horizontal bar marks the calculated range of variation of $\Delta \mu$ given by the formation enthalpy of the compound ($-\Delta H_f \leq \Delta\mu \leq +\Delta H_f$, with $\Delta H_f^\text{calc}=-0.28\,\eV$).
  }
  \label{fig:dconc_defects_dmu}
\end{figure*}

The defect concentrations calculated using the band gap corrected formation energies are shown in \fig{fig:defectstypes} for a representative temperature of 1300\,K ($T_\text{melt}^\AlSb$ = 1327\,K), with the line thicknesses indicating the charge state $|q|$. Figure~\ref{fig:dconc_defects_dmu} shows the dependence of the total defect and net electron concentrations on the chemical environment (chemical potential difference $\Delta\mu$) for a variety of temperatures. For all cases shown here, the intrinsic (self-consistent) electron chemical potential is located near the middle of the gap, although there are slightly more electrons
than holes in the material (intrinsically \ntype material).

\section{Discussion}
\label{sect:discussion}

An inspection of the $G_0W_0$-corrected results in the lower panels of \fig{fig:eform} shows that four different defects have the lowest formation energy and are thus the most abundant, depending on the values of the electron chemical potential, $\mu_e$, and the chemical potential difference, $\Delta\mu$. The aluminum tetrahedral interstitial $\Al_{i,\Al}^{1+}$ is the dominant defect for $\mu_e$ in the lower half of the band gap ($p$-type material), while the aluminum vacancy $V_{\Al}^{3-}$ is dominant for $\mu_e$ in the upper half of the band gap ($n$-type material). For Al-rich conditions (left panel in \fig{fig:eform}), the AlSb antisite defect can also be important when the material is lightly doped $n$-type (near the crossing with $V_{\Al}^{3-}$), particularly considering the uncertainty in the range of $\Delta\mu^{\text{calc}}$ vs $\Delta\mu^{\text{expt}}$ and the possible errors in the formation energies of up to about 0.1\,eV from entropic effects. For Sb-rich conditions (right panel in \fig{fig:eform}), the $\Sb_{\Al}$ antisite defect (neutral or positively charged) is important over a wide range of $\mu_e$, from the $p$-type regime to near the middle of the gap. However, we note that for $\mu_e$ near the middle of the gap (intrinsic or compensated material), the dominant charged native defect is always $\Al_{i,\Al}^{1+}$, since the relevant $\Sb_{\Al}^0$ defect is uncharged.
                                                                
In the absence of impurities, the intrinsic electron chemical potential is located near the middle of the band gap (see \sect{sect:conc}). Under this condition, the two relevant defects are therefore the $\Al_{i,\Al}^{1+}$ interstitial and the $\Sb_{\Al}$ antisite, as illustrated clearly in \fig{fig:defectstypes}. By comparison with \fig{fig:defectsdmu}, we see that the transition point between $\Al_{i,\Al}^{1+}$ and $\Sb_{\Al}^0$ that occurs close to $\Delta\mu=0\,\eV$ corresponds to the minimum in the total defect concentration. In \fig{fig:defectsdmu}, the dominant defect on the Sb-rich (left) side is $\Sb_{\Al}^0$, while on the Al-rich (right) side it is $\Al_{i,\Al}^{1+}$. Near the transition (minimum in the curve), the two concentrations are comparable. In contrast, the minimum in the net electron concentration in \fig{fig:dconcdmu} is located on the Sb-rich side, corresponding to the crossing of the concentrations of $\Al_{i,\Al}^{1+}$ and $\Sb_{\Al}^{1+}$ in \fig{fig:defectstypes}. Since the charge neutrality condition [\eq{eq:chgneutral}] depends only on the concentrations of charged defects, the neutral $\Sb_\Al^0$ antisites do not contribute to charge compensation or net electron concentration.

As mentioned in \sect{sect:conc}, the calculated electron concentration slightly exceeds the hole concentration for pure material, yielding \ntype intrinsic material irrespective of the chemical environment characterized by $\Delta\mu$. This behavior results because the formation energies of acceptor-type defects ($V_{\Al}^{3-}$) always exceed the formation energies of donor-type defects ($\Al_{i,\Al}^{1+}$, $\Sb_{\Al}^{1+}$) when the electron chemical potential is close to the middle of the band gap (see \fig{fig:eform}). It should be noted that this situation is changed if the material is extrinsically doped. For \textit{p}-doped material, the donor-type native defects remain dominant and partially compensate the extrinsic dopant. For \textit{n}-doped material, the ordering of the formation energies is reversed, but the then-dominant acceptor-type native defects still partially compensate the extrinsic dopant.

The temperatures indicated in Figs.~\ref{fig:defectstypes} and \ref{fig:dconc_defects_dmu} refer to thermal equilibrium conditions at those temperatures. In practice, these temperatures can be interpreted as corresponding to annealing temperatures, with the chemical potential difference referring to the chemical environment of the annealing process (e.g., an Sb overpressure corresponds to Sb-rich conditions; conversely, growth is often performed under Al-rich conditions). The highest temperature considered here, 1300\,K, is just below the melting point of 1327\,K and might represent melt growth conditions. However, growth is typically performed too rapidly to allow equilibrium to be achieved and the non-equilibrium grown-in defect concentrations will be higher than predicted here. The curves in \fig{fig:dconc_defects_dmu} essentially represent the predicted concentrations for infinitely long anneals at the specified temperatures.

At 1300\,K, the calculated net electron concentration varies roughly between $10^{16}$ and $10^{17}\,\cm^{-3}$ depending on the chemical potential difference, $\Delta\mu$. If all defects are assumed to be sufficiently mobile down to 700\,K so that the material can reach thermal equilibrium at that temperature, then the lowermost curves in \fig{fig:dconc_defects_dmu} predict a net electron concentration of $10^{10}$ to $10^{11}\,\cm^{-3}$ and a total defect concentration between $10^{10}$ and $10^{12}\,\cm^{-3}$. Since the diffusivity depends exponentially on the inverse temperature, the defect mobilities decrease sharply with temperature. Therefore, as the temperature is further lowered the system will no longer be able to reach the equilibrium concentration in reasonable time, which requires excess defects either to diffuse to the surface or to anneal by recombination. In contrast to lattice defects, the intrinsic electron and hole concentrations, $n_e$ and $n_h$, readily adjust to temperature changes. The ``freezing in'' of the defect concentrations is therefore expected to crucially affect the charge neutrality condition upon cooling to low (e.g., room) temperatures (in particular if there are no extrinsic dopants). However, since the diffusivities of the individual defects are currently unknown, we cannot quantitatively describe this ``freezing in'' of the defect distributions in the present study; thus, predictions of the charge carrier and defect concentrations near room temperature and below may be unreliable. The determination of diffusivities for specific defects, to account for the kinetics of defect ``freeze-in,'' will be the subject of future work.

Experimentally, AlSb crystals grown from the melt have often been found to display \ptype conductivity.\cite{KJ_pers_comm} The present finding that the pure material behaves intrinsically \ntype is, however, not in contradiction with this observation. A typical experimental setup employs a graphite susceptor, alumina crucible, and quartz tubes for melting Sb, which are potential sources of various impurities, most importantly carbon, oxygen, silicon, and aluminum.\cite{KJ_pers_comm} In particular, carbon impurities act as acceptors, therefore accidental \ptype doping is a very likely scenario which is supported by chemical analysis of grown AlSb crystals and measurements on intentionally doped samples.\cite{KJ_pers_comm} The measured levels of carbon impurities and the experimental observation that the material becomes intrinsic around 1000\,K (Ref.~[\onlinecite{MadelungSemiHandbook}]) are consistent with our calculated excess native electron concentration at that temperature. A detailed study of the role of extrinsic defects in AlSb is beyond the scope of the present work; however, the results of ongoing work to elucidate the effects of impurities and to investigate ways to optimize the electronic properties of the material by intentional doping are forthcoming.

\section{Conclusions}

In summary, we have employed density functional theory calculations to study the properties of intrinsic point defects in aluminum antimonide. An exhaustive set of defect configurations ---including vacancies, antisites, and interstitials (tetrahedral, hexagonal, and split), with all relevant charge states--- was considered based on knowledge of other III-V compounds. Relaxed atomic structures of each defect were carefully determined, and formation energies were calculated to evaluate the equilibrium concentrations of each defect. Strain and electrostatic artifacts related to the use of the supercell approach were carefully removed by employing a finite-size scaling procedure, which involved performing a series of calculations for each defect with supercell sizes ranging from 32 to 216 atoms. The underestimation of the band gap due to the local density approximation was taken into account by applying an {\it a posteriori} correction scheme that utilized separate {\it GW} calculations to obtain valence and conduction band offsets, which enter the correction.

Aluminum interstitials ($\Al_{i,\Al}^{1+}$), antimony antisites ($\Sb_{\Al}^{0}$, $\Sb_{\Al}^{1+}$), and aluminum vacancies ($V_{\Al}^{3-}$) were found to be the most dominant defects, depending on the electron chemical potential and the chemical potential difference (chemical environment) of the system. We observe that $\Al_{i,\Al}^{1+}$ interstitials and $\Sb_{\Al}$ antisites dominate under Al-rich and Sb-rich conditions, respectively. Calculated formation energies were employed in solving the charge neutrality condition to obtain self-consistent defect concentrations and intrinsic electron chemical potential for the pure material at various temperatures. We find the material to be intrinsically weakly \ntype\ and predict both the total defect and the net electron concentrations. Near the melting point, the equilibrium concentration of native defects is predicted to be in the $10^{16}$ to $10^{17}$\,cm$^{-3}$ range, while at lower temperatures, it is expected that concentrations down to the $10^{10}$\,cm$^{-3}$ range or lower can be achieved. The net excess electron density in bulk grown material might be as high as $10^{9}$ to $10^{11}$\,cm$^{-3}$ from ``freeze-in'' of defects from melt solidification.

For extrinsically doped material, which we do not treat explicitly in detail in this work, the dominant native defects depend on the nature of the doping. For \textit{n}-doped material, $V_{\Al}^{3-}$ and $\Al_\Sb^{2-}$ tend to be important, while for \textit{p}-doped material, $\Al_{i,\Al}^{1+}$ and $\Sb_\Al^{1+}$ are important, depending on chemical environment (Al-rich vs. Sb-rich). Some amount of self-compensation from the native defects occurs in both cases.

Finally, we note that the present work is part of a concerted research effort which ultimately aims to provide a complete and consistent picture of the point defect properties of AlSb and the relations to carrier transport properties. We are engaged in further theoretical and experimental work to explore the role of extrinsic defects, as well as the scattering behavior of defects on carrier transport. In this context, the present study forms the basis for these future studies, which will be the subjects of forthcoming reports.

\begin{acknowledgments}
This work was performed under the auspices of the U.S. Department of Energy by Lawrence Livermore National Laboratory in part under Contract W-7405-Eng-48 and in part under Contract DE-AC52-07NA27344. The authors acknowledge support from the National Nuclear Security Administration Office of Nonproliferation Research and Development (NA-22) and from the Laboratory Directed Research and Development Program at LLNL. The authors would like to thank Babak Sadigh for fruitful discussions.
\end{acknowledgments}

\appendix*{}
\section{Band gap correction of defect formation energies}

While DFT calculations typically give very reasonable values for energy differences {\em within} a group of bands (e.g., the valence band), energy differences {\em between} different groups of bands (i.e., band gaps, e.g., between the valence and conduction bands) are much less reliable. This shortcoming particularly affects the differences between the valence and conduction bands, giving rise to the well-known band gap error.

The incorrect description of the energy differences between different groups of bands can affect the total energy. The most sensitive contribution is the band energy which is given by
\begin{align}
  E_b &= \sum_i \sum_k f_{ik} \epsilon_{ik},
\end{align}
where $i$ and $k$ run over bands and $k$-points, respectively, and $f_{ik}$ and $\epsilon_{ik}$ are the occupation numbers and eigenvalues, respectively. Without loss of generality, one can divide the band energy into separate sums over the valence and conduction band states, as
\begin{align}
  E_b &=
  \sum_i^{\VB}   \sum_k f_{ik} \epsilon_{ik}
  + \sum_i^{\CB} \sum_k f_{ik} \epsilon_{ik}.
  \label{eq:band1}
\end{align}
If one assumes rigid levels, which is a reasonable approximation in many cases, the errors in the energy differences between two groups of bands (i.e., across a band gap) can be corrected by adding constant energy shifts to the valence and conduction band states:
\begin{align*}
  \epsilon_{ik} \rightarrow \epsilon_{ik}+\Delta E_{\VB}
  & \qquad \text{for the valence band, and} \\
  \epsilon_{ik} \rightarrow \epsilon_{ik}+\Delta E_{\CB}
  & \qquad \text{for the conduction band}.
\end{align*}
The sum of the band shifts, $\Delta E_{\VB} + \Delta E_{\CB}$, equals the band gap error. The expression for the corrected band energy then reads
\begin{align}
  \widetilde{E}_b
  &=
  \sum_i^{\VB}   \sum_k f_{ik} (\epsilon_{ik}+\Delta E_{\VB})
  \nonumber
  \\
  &\quad
  + \sum_i^{\CB} \sum_k f_{ik} (\epsilon_{ik}+\Delta E_{\CB}).
  \label{eq:band2}
\end{align}
Taking the difference between Eqs.~\eqref{eq:band1} and
\eqref{eq:band2}, one obtains the band energy correction term
\begin{align}
  \Delta E_b^{corr}
  &=\left(\widetilde{E}_b - E_b\right)_\text{defect} 
           -\left(\widetilde{E}_b - E_b\right)_\text{ideal}
     \nonumber
     \\
  &= \Delta E_{\VB} \underbrace{\sum_i^{\VB}
   \sum_k \left(f_{ik}-1\right)}_{\displaystyle \Delta z_h}
   +  \Delta E_{\CB} \underbrace{\sum_i^{\CB}
   \sum_k f_{ik},}_{\displaystyle \Delta z_e}
   \label{eq:z_eh}
\end{align}
where $\Delta z_h$ is simply the number of unoccupied states in the valence band and $\Delta z_e$ is the number of occupied states in the conduction band.

According to \eq{eq:eform}, the defect formation energy of charged defects further depends on the position of the valence band maximum. The total correction term for the formation energy thus reads
\begin{align}
  \Delta E^{corr}
  = (q + \Delta z_h) \Delta E_{\VB} + \Delta z_e \Delta E_{\CB}.
  \tag{\ref{eq:EG_corr}}
\end{align}

The energy offsets $\Delta E_{\VB}$ and $\Delta E_{\CB}$ can be obtained, for example, from \textit{GW} calculations, which for many systems provide band gaps and structures in good agreement with experiment.

It is important to realize that this scheme neglects both level relaxations and changes in the double counting term. If these limitations are acceptable, this method offers a simple {\it a posteriori} correction of the formation energies. It should furthermore be noted that transition levels are not affected by the relative weight of $\Delta E_{\VB}$ and  $\Delta E_{\CB}$. Upon application of this correction scheme, acceptor levels track the valence band maximum while donor transitions follow the conduction band minimum.

\end{document}